\documentclass[useAMS,usenatbib]{mn2e}
\newcommand{\bib}{\bibitem[\protect\citeauthoryear}
\usepackage[dvips]{graphicx}
\usepackage[T1]{fontenc}
\usepackage{ae,aecompl}

\title[3C\,296 jet model]{A relativistic model of the radio jets in 3C\,296}
\author[R.A. Laing et al.]{R.A. Laing\thanks{E-mail: rlaing@eso.org}$^{1}$,
J.R. Canvin$^{2,3}$, A.H.Bridle$^{4}$, M.J. Hardcastle$^{5}$\\$^1$ European
Southern Observatory, Karl-Schwarzschild-Stra\ss e 2, D-85748
Garching-bei-M\"{u}nchen, Germany \\$^2$ School of Physics, University of
Sydney, A28, Sydney, NSW 2006, Australia\\$^3$ University of Oxford, Department
of Astrophysics, Denys Wilkinson Building, Keble Road, Oxford OX1 3RH \\$^4$
National Radio Astronomy Observatory, 520 Edgemont Road, Charlottesville, VA
22903-2475, U.S.A.\\$^5$ School of Physics, Astronomy and Mathematics,
University of Hertfordshire, College Lane, Hatfield, Hertfordshire AL10 9AB}

\date{Received}

\pagerange{\pageref{firstpage}--\pageref{lastpage}} \pubyear{2006}

\begin{document}
\label{firstpage}

\maketitle

\begin{abstract}
We present new, deep 8.5-GHz VLA observations of the nearby, low-luminosity
radio galaxy 3C\,296 at resolutions from 0.25 to 5.5\,arcsec.  These show the
intensity and polarization structures of the twin radio jets in detail. We
derive the spectral-index distribution using lower-frequency VLA observations
and show that the flatter-spectrum jets are surrounded by a sheath of
steeper-spectrum diffuse emission. We also show images of Faraday rotation
measure and depolarization and derive the apparent magnetic-field structure.  We
apply our intrinsically symmetrical, decelerating relativistic jet model to the
new observations. An optimized model accurately fits the data in both total
intensity and linear polarization. We infer that the jets are inclined by
$58^\circ$ to the line of sight.  Their outer isophotes flare to a half-opening
angle of 26$^\circ$ and then recollimate to form a conical flow beyond 16\,kpc
from the nucleus.  On-axis, they decelerate from a (poorly-constrained) initial
velocity $\beta = v/c \approx 0.8$ to $\beta \approx 0.4$ around 5\,kpc from the
nucleus, the velocity thereafter remaining constant. The speed at the edge of
the jet is low everywhere. The longitudinal profile of proper emissivity has
three principal power-law sections: an inner region (0 -- 1.8\,kpc), where the
jets are faint, a bright region (1.8 -- 8.9\,kpc) and an outer region with a
flatter slope.  The emission is centre-brightened.  Our observations rule out a
globally-ordered, helical magnetic-field configuration. Instead, we model the
field as random on small scales but anisotropic, with toroidal and longitudinal
components only. The ratio of longitudinal to toroidal field falls with distance
along the jet, qualitatively but not quantitatively as expected from flux
freezing, so that the field is predominantly toroidal far from the nucleus.  The
toroidal component is relatively stronger at the edges of the jet.  A simple
adiabatic model fits the emissivity evolution only in the outer region after the
jets have decelerated and recollimated; closer to the nucleus, it predicts far
too steep an emissivity decline with distance. We also interpret the
morphological differences between brightness enhancements (``arcs'') in the main
and counter-jets as an effect of relativistic aberration.
\end{abstract}


\section{Introduction}
\label{intro}

The case that the jets in low-luminosity, FR\,I radio galaxies \citep{FR74} are
initially relativistic and decelerate to sub-relativistic speeds on kiloparsec
scales rests on several independent lines of evidence, as follows.
\begin{enumerate}
\item Proper motions corresponding to speeds comparable with or exceeding $c$
  have been observed directly on pc and kpc scales in FR\,I jets
  \citep{Giov01,Biretta95,Hard03}.
\item FR\,I sources must be the side-on counterparts of at least a subset of the
  BL Lac population, in which evidence for bulk relativistic flow on pc scales
  is well established (e.g. \citealt{UP95}).
\item FR\,I jets show side-to-side asymmetries which decrease with distance from
  the nucleus, most naturally explained by Doppler beaming of emission from an
  intrinsically symmetrical, decelerating, relativistic flow \citep{LPdRF}, in
  which case the brighter jet is approaching us.  There is continuity of
  sidedness from pc to kpc scales and superluminal motion, if observed, is in
  the brighter jet.
\item The lobe containing the brighter jet is less depolarized than the
  counter-jet lobe, consistent with Faraday rotation in the surrounding halo 
  of hot plasma if the brighter jet is approaching \citep{L88,Morganti97}.
\item Theoretical work has demonstrated that initially relativistic jets can
  decelerate from relativistic to sub-relativistic speeds without disruption
  provided that they are not too powerful, the surrounding halo of hot plasma
  providing a pressure gradient large enough to recollimate them 
  \citep{Phi83,Bic94,Kom94,BLK96}.
\end{enumerate}
We have developed a method of modelling jets on the assumption that they are
intrinsically symmetrical, axisymmetric, relativistic, decelerating flows. By
fitting to deep radio images in total intensity and linear polarization, we can
determine the three-dimensional variations of velocity, emissivity and
magnetic-field ordering.  We first applied the technique to the jets in the
radio galaxy 3C\,31 \citep[LB]{LB02a}. With minor revisions, the same method
proved capable of modelling the jets in B2\,0326+39 and B2\,1553+24
\citep[CL]{CL} and in the giant radio galaxy NGC\,315 \citep[CLBC]{CLBC}. We
used the derived velocity field for 3C\,31, together with {\em Chandra}
observations of the hot plasma surrounding the parent galaxy, to derive the run
of pressure, density, Mach number and entrainment rate along its jets via a
conservation-law analysis \citep{LB02b}. We also examined adiabatic models for
3C\,31 in detail \citep{LB04}.

We now present a model for the jets in 3C\,296, derived by fitting to new,
sensitive VLA observations at 8.5\,GHz.  In Section~\ref{obs}, we introduce
3C\,296 and summarize our VLA observations and their reduction. Images of the
source at a variety of resolutions are presented in Section~\ref{descrip}, where
we also describe the distributions of spectral index, rotation measure and
apparent magnetic field derived by combining our new images with lower-frequency
data from the VLA archive. We briefly recapitulate our modelling technique in
Section~\ref{model} and compare the observed and model brightness and
polarization distributions in Section~\ref{results}. The derived geometry,
velocity, emissivity and field distributions are presented in
Section~\ref{physical}.  We compare the result of model fitting for the five
sources we have studied so far in Section~\ref{sourcecomp} and summarize our
results in Section~\ref{summary}.

We adopt a concordance cosmology with Hubble constant, $H_0$ =
70\,$\rm{km\,s^{-1}\,Mpc^{-1}}$, $\Omega_\Lambda = 0.7$ and $\Omega_M =
0.3$. Spectral index, $\alpha$, is defined in the sense $S(\nu) \propto
\nu^{-\alpha}$ and we use the notations $P = (Q^2+U^2)^{1/2}$ for polarized
intensity and $p = P/I$ for the degree of linear polarization. We also define
$\beta = v/c$, where $v$ is the flow velocity. $\Gamma = (1-\beta^2)^{-1/2}$ is
the bulk Lorentz factor and $\theta$ is the angle between the jet axis and the
line of sight.

\section{Observations and images}
\label{obs}

\subsection{3C\,296}
\label{3c296}

For our modelling technique to work, both radio jets must be
detectable at high signal-to-noise ratio in linear polarization as well as
total intensity, separable from any surrounding lobe emission, straight and
antiparallel.  3C\,296 satisfies all of these criteria, although the lobe
emission is relatively brighter than in the other sources we have studied. This
is a potential source of error, as we shall discuss.

3C\,296 was first imaged at 1.4 and 2.7\,GHz by \citet{BLP}. VLA observations
showing the large-scale structure at 1.5\,GHz and the inner jets at 8.5\,GHz
were presented by \citet{LP91} and \citet{Hard97} respectively. 

The large-scale radio structure of 3C\,296 has two radio lobes with well-defined
outer boundaries and weak diffuse ``bridge'' emission around much of the path of
both jets, at least in projection \citep{LP91}.  3C\,296 is therefore more
typical of a ``bridged twin-jet'' FR\,I structure \citep{L96} than the ``tailed
twin-jet'' type exemplified by 3C\,31 and other sources we have studied so far.
\citet{PDF96} point out that the ``lobed, bridged'' FR\,I population is more
abundant in a complete sample of low-luminosity radio galaxies than the
``plumed, tailed'' population.  It is therefore particularly interesting to
compare the inferred properties of the jets in 3C\,296 with those of the sources
we have modelled previously.  Of the two lobes, that containing the counter-jet
depolarizes more rapidly with decreasing frequency between 1.7 and 0.6\,GHz
\citep*{GHS}, consistent with the idea that it is further away from us.

The parsec-scale structure, imaged by \citet{Giov05}, is two sided, but brighter
on the same side as the kpc-scale jet emission.  Non-thermal X-ray and
ultraviolet emission have been detected from the brighter jet \citep{Hard05} and
the nucleus \citep{CCC,Hard05}. The source is associated with the giant
elliptical galaxy NGC\,5532 ($z = 0.02470 \pm 0.00007$), the central, dominant
member of a small group \citep{Miller02}. The linear scale is
0.498\,kpc\,arcsec$^{-1}$ for our adopted cosmology.  The galaxy has a slightly
warped nuclear dust ellipse in PA $163^\circ \pm 4^\circ$ with an inclination
of $73^\circ \pm 4^\circ$ assuming intrinsic circularity \citep[parameters are
taken from the last reference]{Martel99,deKoff00,VKdZ05}.  X-ray emission from
hot plasma associated with the galaxy was detected by {\em ROSAT} and imaged by
{\em Chandra} \citep{Miller99,HW99,Hard05}.

\subsection{Observations}

New, deep VLA observations of 3C\,296 at 8.5\,GHz are presented here. We were
interested primarily in the inner jets, and therefore used a single pointing
centre at the position of the nucleus and the maximum bandwidth (50\,MHz) in each
of two adjacent frequency channels. In order to cover the full range of spatial
scales in the jets at high resolution, we observed in all four VLA
configurations.  We combined our new datasets with the shorter observations
described by \citet{Hard97}, which were taken with the same array
configurations, centre frequency and bandwidth.

In order to correct the observed ${\bf E}$-vector position angles at 8.5\,GHz
for the effects of Faraday rotation and to estimate the spectrum of the source,
we also reprocessed earlier observations at 4.9\,GHz and 1.4 -- 1.5\,GHz
(L-band).  The 4.9-GHz data were taken in the A configuration. At L-band, we
used two A-configuration observations (the later one also from \citealt{Hard97})
and the combined B, C and D-configuration dataset from \citet{LP91}.  A journal
of observations is given in Table~\ref{tab:journal}.

\begin{center}
\begin{table}
\caption{Journal of VLA observations. $\nu$ and $\Delta\nu$ are the centre
  frequencies and bandwidth, respectively, for the one or two frequency channels
  observed and t is the on-source integration time scaled to an array with all
  27 antennas operational. References to previous publications using the same
  data: 1 \citet{Hard97}, 2 \citet{LP91}.}
\begin{center}
\begin{tabular}{clllcl}
\hline
 Config- &    Date     & $\nu$ & $\Delta\nu$  & t & Ref\\
 uration &             & (GHz)     & (MHz)      & (min)  & \\
\hline
        A  & 1995 Jul 24 & 8.4351, 8.4851   & 50 & 123 & 1 \\
        B  & 1995 Nov 27 & 8.4351, 8.4851   & 50 & 110 & 1 \\
        C  & 1994 Nov 11 & 8.4351, 8.4851   & 50 &  47 & 1 \\
        D  & 1995 Mar 06 & 8.4351, 8.4851   & 50 &  30 & 1 \\
        A  & 2004 Dec 12 & 8.4351, 8.4851   & 50 & 250 &   \\
        A  & 2004 Dec 13 & 8.4351, 8.4851   & 50 & 253 &   \\
        B  & 2005 Apr 21 & 8.4351, 8.4851   & 50 & 250 &   \\
        B  & 2005 Apr 23 & 8.4351, 8.4851   & 50 & 264 &   \\
        C  & 2005 Jul 14 & 8.4351, 8.4851   & 50 & 249 &   \\
        D  & 2005 Nov 27 & 8.4351, 8.4851   & 50 &  98 &   \\
        A  & 1981 Feb 13 & 4.8851           & 50 & 130 &   \\
        A  & 1981 Feb 13 & 1.40675          &12.5& 125 &   \\
        A  & 1995 Jul 24 & 1.3851, 1.4649   & 50 &  45 & 1 \\
        B  & 1987 Dec 08 & 1.4524, 1.5024   & 25 &  60 & 2 \\
        C  & 1988 Mar 10 & 1.4524, 1.5024   & 25 &  50 & 2 \\
        D  & 1988 Oct  7 & 1.4524, 1.5024   & 25 &  80 & 2 \\
\hline 
\end{tabular}
\end{center}
\label{tab:journal}
\end{table}
\end{center}

\subsection{Data reduction}

\subsubsection{8.5\,GHz observations}

All of the 8.5-GHz datasets were calibrated using standard procedures in the
{\sc aips} package. 3C\,286 was observed as a primary amplitude calibrator and
to set the phase difference between right and left circular polarizations;
J1415+133 was used as a secondary phase and amplitude calibrator and to
determine the instrumental polarization. After initial calibration, the 1994-5
data were precessed to equinox J2000 and shifted to the phase centre of the new
observations.  All three A-configuration datasets were then imaged to determine
a best estimate for the position of the core (RA 14 16 52.951; Dec. 10 48
26.696; J2000) and then self-calibrated, starting with a phase-only solution for
a point-source model at the core position and performing one further phase and
one amplitude-and-phase iteration with {\sc clean} models. The core varied
significantly between observations, so we added a point source at the position
of the peak to the 1995 dataset to equalize the core flux densities. We then
combined the three A-configuration datasets and performed one further iteration
of phase self-calibration. The B-configuration datasets were initially phase
self-calibrated using the A-configuration model. After further iterations of
self-calibration and adjustment of the core flux density, they were concatenated
with the A-configuration dataset. The process was continued to add the C- and
D-configuration datasets, with relative weights chosen to maintain a resolution
of 0.25\,arcsec for images from the final combination.

The off-source noise levels for the final $I$ images ($\approx 4\mu$Jy
beam$^{-1}$ rms at full resolution) are close to the limits set by thermal
noise, although there are slight residual artefacts within a pair of triangles
centred on the core and defined by position angles $\pm15^\circ$.  These do not
affect the jet emission, and the maximum increase in noise level is in any case
only to $\approx 6\mu$Jy beam$^{-1}$ rms at full resolution.  $I$ images were
made using {\sc clean} and maximum-entropy deconvolution, the latter being
convolved with the same Gaussian beam used to restore the {\sc clean}
images. Differences between the two deconvolutions were very subtle. As usual,
the maximum-entropy method gave smoother images, avoiding the mottled appearance
produced by {\sc clean} in regions of uniform surface brightness. On the other
hand, the {\sc clean} images had slightly lower off-source noise levels,
significantly lower (and effectively negligible) zero-level offsets, and
improved fidelity in regions with high brightness gradients. The differences
between the two methods are sufficiently small that none of our quantitative
conclusions is affected by choice of method. We have used the {\sc clean}
deconvolutions for all of the modelling described in Sections~\ref{model} --
\ref{physical}, but have generally preferred the maximum entropy images for the
grey-scale displays in Section~\ref{descrip}.  All $Q$ and $U$ images were {\sc
clean}ed.

We made $I$, $Q$ and $U$ images at standard resolutions of 5.5, 1.5, 0.75 and
0.25\,arcsec FWHM.  All images were corrected for the primary-beam response
during or after deconvolution, so the noise levels (listed in
Table~\ref{tab:noise}) strictly apply only in the centre of the field.  The
total flux density determined by integrating over a {\sc clean} image at 5.5\,arcsec
resolution is 1.31\,Jy (with an estimated error of 0.04\,Jy due to calibration
uncertainties alone), compared with an expected value of $1.20 \pm 0.06$\,Jy,
interpolated between single-dish measurements at 10.7 and 5.0\,GHz
\citep{LP80}. At least in the low-resolution images, we have therefore recovered
the total flux density of the source, despite the marginal spatial sampling
(Table~\ref{tab:noise}).

\begin{center}
\begin{table}
\caption{Resolutions and noise levels for the images used in this
  paper. $\sigma_I$ is the off-source noise level on the $I$ image; $\sigma_P$
  the average of the noise levels for $Q$ and $U$. M and C denote
  maximum-entropy and CLEAN $I$ deconvolutions.  All noise levels were
  determined before correction for the primary beam response and larger values
  are appropriate at significant distances from the field centre. The
  approximate maximum scale of structure imaged reliably is also given
  \citep{ObsSS}. Note that the 1.5-arcsec L-band image includes A-configuration
  data at 1.41\,GHz and BCD configuration data at 1.45/1.50\,GHz (see text).
\label{tab:noise}}
\begin{tabular}{lllcccc}
\hline
 FWHM     & Freq & Config-  &\multicolumn{3}{c}{rms noise level} & Max \\
  (arcsec)& GHz  & urations &\multicolumn{3}{c}{[$\mu$Jy\,beam$^{-1}$]}& scale \\
          &      &          &\multicolumn{2}{c}{$\sigma_I$}& $\sigma_P$ & arcsec\\
          &      &          &  M & C & &\\
\hline
5.5 & 8.4601  &ABCD& $-$  & 15   & 8.8&  180 \\
5.5 & 1.5024  & BCD& $-$  &  40  & 40  &  900 \\
5.5 & 1.4524  & BCD & $-$  & 90   & 35  &  900 \\
5.5 & 1.4774  & BCD& $-$  & 38   & $-$ &  900 \\
1.5 & 8.4601  &ABCD& 9.5  & 9.1  & 4.8 &  180 \\
1.5 & 1.3851  &A   & $-$  & $-$  & 37  &  40 \\
1.5 & 1.4649  &A   & $-$  & $-$  & 35  &  40 \\
1.5 & 1.40675 &ABCD&  27  & $-$   & $-$ &  900 \\
0.75 & 8.4601 &ABCD& 4.3  & 4.2  & 4.1& 180  \\
0.40 & 8.4601 &ABCD& $-$  & 4.2  & $-$& 180  \\
0.40 & 4.8851 &A   & $-$  & 33   & $-$&  10 \\
0.25 & 8.4601 &ABCD& 4.9  & 3.6  & 4.5  & 180  \\
\hline
\end{tabular}
\end{table}
\end{center}

\subsubsection{4.9 and 1.4--1.5\,GHz observations}

For the B, C and D-configuration observations at 1.45 and 1.50\,GHz, we started
from the combined dataset described by \citet{LP91}. We imaged a significant
fraction of the primary beam using a faceting procedure in order to minimize
errors due to non-coplanar baselines and confusion, using the resulting {\sc
clean} model as the input for a further iteration of phase-only
self-calibration. This led to a significant improvement in off-source noise
level compared with the values obtained by \citealt{LP91} using earlier
algorithms. The noise level in $I$ is much larger at 1.45\,GHz than at
1.50\,GHz, probably as a result of interference; the $Q$ and $U$ images in the
two frequency channels have similar noise levels, however.  We used both
single-frequency and combined images at 5.5-arcsec resolution; the former for
polarization and the latter for spectral index. The reason was that the
variation of Faraday rotation across the structure turned out to be large enough
that the ${\bf E}$-vector position angle difference between the two frequency
channels changed significantly across the structure (Section~\ref{RM}), whereas
the variations of spectral index did not lead to detectable differences between
the $I$ images in the two channels (Section~\ref{spectrum}).

The A-configuration observations at 1.39, 1.41, 1.46 and 4.9\,GHz were
calibrated, self-calibrated and imaged using standard methods.  The 1.41-GHz A
configuration dataset had better spatial coverage (in multiple separated
snapshots) but poorer sensitivity compared with the later short observations at
1.39/1.46\,GHz. We were able to make a satisfactory multi-configuration $I$
image at 1.5 arcsec resolution using the 1.41-GHz dataset together with the
1.45/1.50-GHz BCD configuration combination. No core variability was apparent
between observations. The frequency difference is sufficiently small that there
should be no significant effects on the spectral index image of the jets (see
Section~\ref{spectrum}). The {\sc clean} deconvolution of this combined ABCD
configuration L-band image image had significant artefacts, but the
maximum-entropy image at this resolution was adequate, at least in the area of
the jets.  In order to determine rotation measures, however, the individual IFs
were kept separate (exactly as in the lower-resolution case) and we therefore
made $Q$ and $U$ images from the individual A-configuration datasets.

The A-configuration image at 4.9\,GHz shows only the innermost regions of the
jets. No significant polarization is detected and we use it only to estimate the
spectral index of the jets at high resolution.

Details of the images are again given in Table~\ref{tab:noise}.


\section{Total intensity, spectrum, Faraday rotation and apparent magnetic field}
\label{descrip}

\subsection{Total-intensity images}
\label{images}

\begin{figure}
\includegraphics[width=8.5cm]{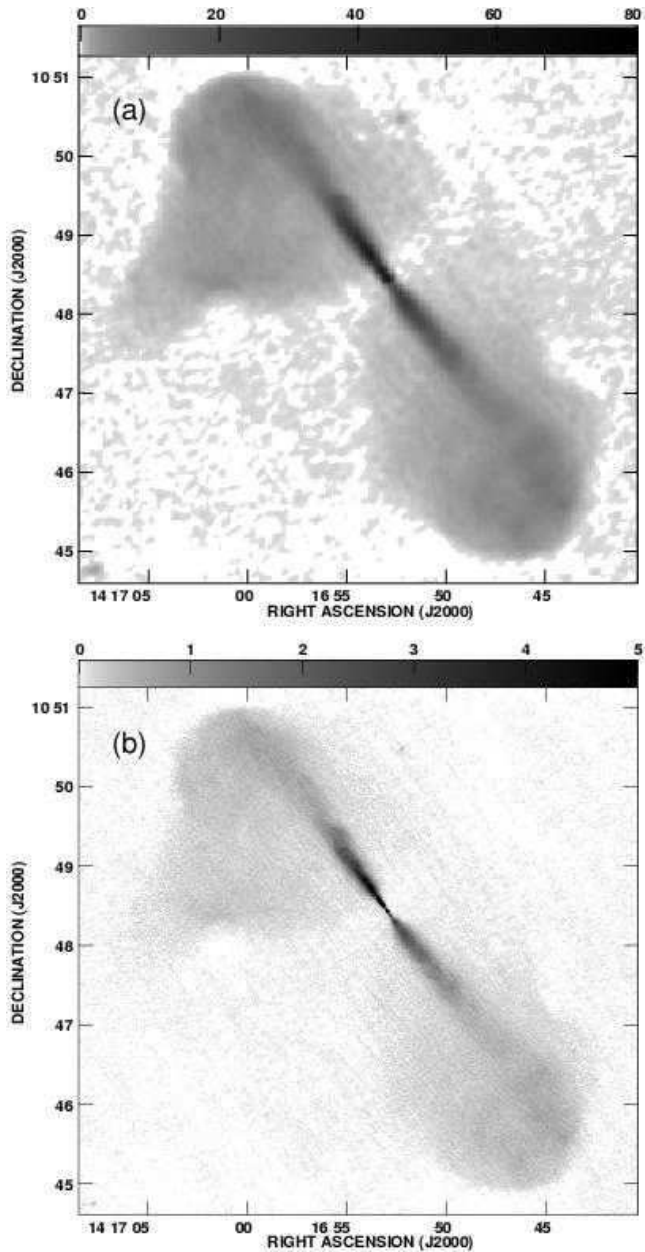}
\caption{Grey-scales of total intensity. A square-root transfer function has
  been used to emphasise the low-brightness emission.  (a) 1.48-GHz BCD CLEAN
  image at 5.5-arcsec FWHM resolution. The grey-scale range, 0 -- 80\,mJy (beam
  area)$^{-1}$, is marked by the labelled wedge. (b) L-band ABCD maximum entropy
  image at 1.5-arcsec FWHM resolution from a combination of 1.41 and 1.48-GHz
  datasets. The grey-scale range is 0 -- 5\,mJy beam$^{-1}$. The areas
  covered by the two panels are identical.
\label{fig:ifull}}
\end{figure}

\begin{figure*}
\includegraphics[width=12cm]{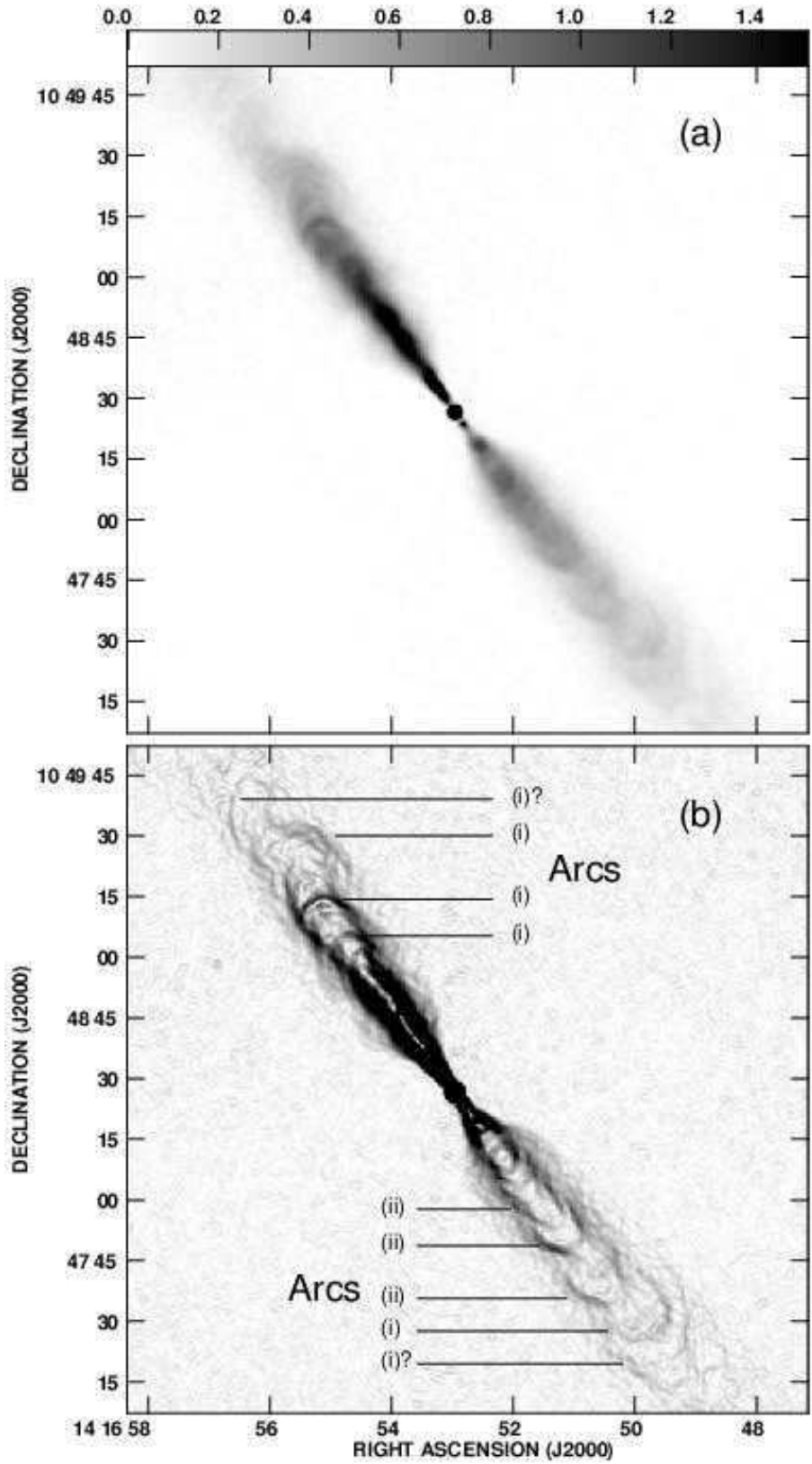}
\caption{(a) Grey-scale of total intensity for a maximum-entropy image of the jets of 3C\,296 at
  8.5\,GHz. The resolution is 1.5\,arcsec FWHM and the grey-scale range is 0 --
  1.5\,mJy beam$^{-1}$. (b) Grey-scale of a Sobel-filtered version of this
  image. Narrow features in the brightness distribution (``arcs'') are marked
  with their types as defined in Section~\ref{images}.
  \label{fig:i1.5}}
\end{figure*}

\begin{figure}
\includegraphics[width=8.5cm]{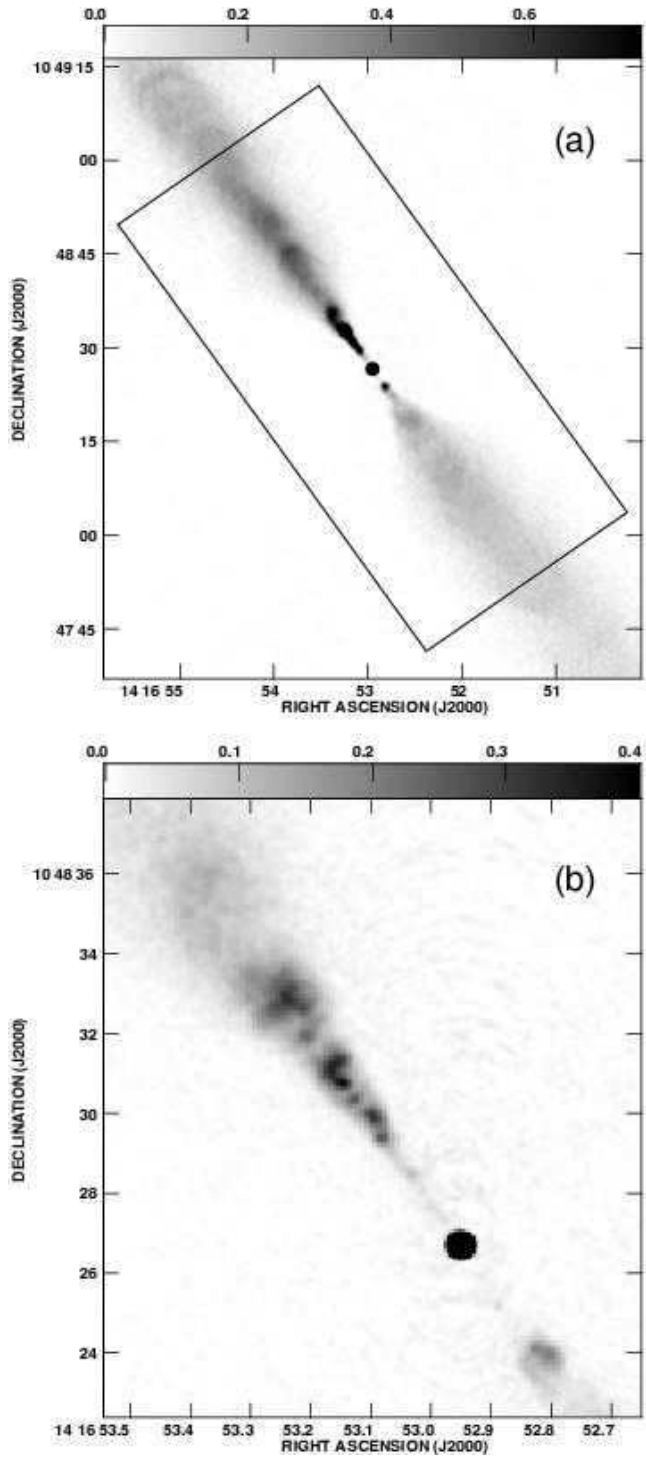}
\caption{Grey-scales of total intensity for the jets of 3C\,296 at 8.5\,GHz,
  both from maximum-entropy images, corrected for attenuation by the primary
  beam. (a) Image at 0.75-arcsec resolution.  The grey-scale range is 0 --
  0.75\,mJy beam$^{-1}$ and the box shows the modelled area
  (Sections~\ref{model} -- \ref{physical}).  (b) Image at 0.25-arcsec resolution
  with a grey-scale range of 0 -- 0.4\,mJy beam$^{-1}$.
\label{fig:ihires}}
\end{figure}

The overall structure of the source at 5.5-arcsec FWHM is shown in
Fig.~\ref{fig:ifull}(a)\footnote{This is essentially the same image as in figs
13 and 30 of \citet{LP91}}. This image emphasises that the lobes are
well-defined, with sharp leading edges. With the possible exception of the
innermost $\pm$15\,arcsec, where the extended emission is very faint, the jets
appear superimposed on the lobes everywhere. It is possible that the jets
propagate (almost) entirely within the lobes, although the superposition may
also be a projection effect. Fig.~\ref{fig:ifull}(b), which shows the same area
at a resolution of 1.5\,arcsec FWHM, emphasises that the boundaries between the
jets and lobes are not precisely defined at 1.4\,GHz, although the edges of the
jets are marked by sharp brightness gradients.

In contrast, the lobe emission is almost invisible in an 8.5-GHz image at the
same resolution (Fig.~\ref{fig:i1.5}a).  The transfer function of this
grey-scale has been chosen to emphasise sub-structure (``arcs'') in the
brightness distributions. These features are emphasised further and labelled in
Fig.~\ref{fig:i1.5}(b), which shows a Sobel-filtered version of the same image.
The Sobel operator \citep{Pratt} computes an approximation to $\mid
\nabla(I)\mid$ and therefore highlights large brightness gradients.  Arcs are
observed in both jets, but appear to have two characteristic shapes: (i) concave
towards the nucleus and approximately semicircular and (ii) oblique to the jet
direction on either side of the jet but without large brightness gradients
on-axis. They do not occupy the full width of the jets. All of the arcs in the
main jet are of type (i); of those in the counter-jet, the inner three marked in
Fig.~\ref{fig:i1.5}(b) are of type (ii) but the outer two are of type (i).  In
Section~\ref{arcs}, we interpret the structural differences in the main and
counter-jet arcs as an effect of relativistic aberration.

Fig.~\ref{fig:ihires} shows the inner jets of 3C\,296 in grey-scale
representations again chosen to emphasise the fine-scale structure. The initial
``flaring'', i.e., decollimation followed by recollimation \citep{B82}, is
clearest at 0.75-arcsec resolution (Fig.~\ref{fig:ihires}a), as is the ridge of
emission along the axis of the main jet noted by \citet{Hard97}; the innermost
arcs are also visible. Our highest-resolution image (0.25\,arcsec FWHM;
Fig.~\ref{fig:ihires}b) shows the faint base of the main jet. This brightens
abruptly at a distance of 2.7 arcsec from the nucleus; further out there is
complex, non-axisymmetric knotty structure within a well-defined outer
envelope. All of these features are typical of FR\,I jet bases (e.g.\ LB, CL,
CLBC). There is only a hint of a knot in the counter-jet at 1.8\,arcsec from the
nucleus, opposite the faint part of the main jet, but integrating over
corresponding regions between 0.5 and 2.7 arcsec in the main and counter-jets
gives $I_{\rm j}/I_{\rm cj} \approx 2.7$. The counter-jet brightens at
essentially the same distance from the nucleus as the main jet, but its
prominent, non-axisymmetric, edge-brightened knot is significantly wider than
the corresponding emission on the main jet side.

\subsection{Spectrum}
\label{spectrum}

\begin{figure*}
\includegraphics[width=15cm]{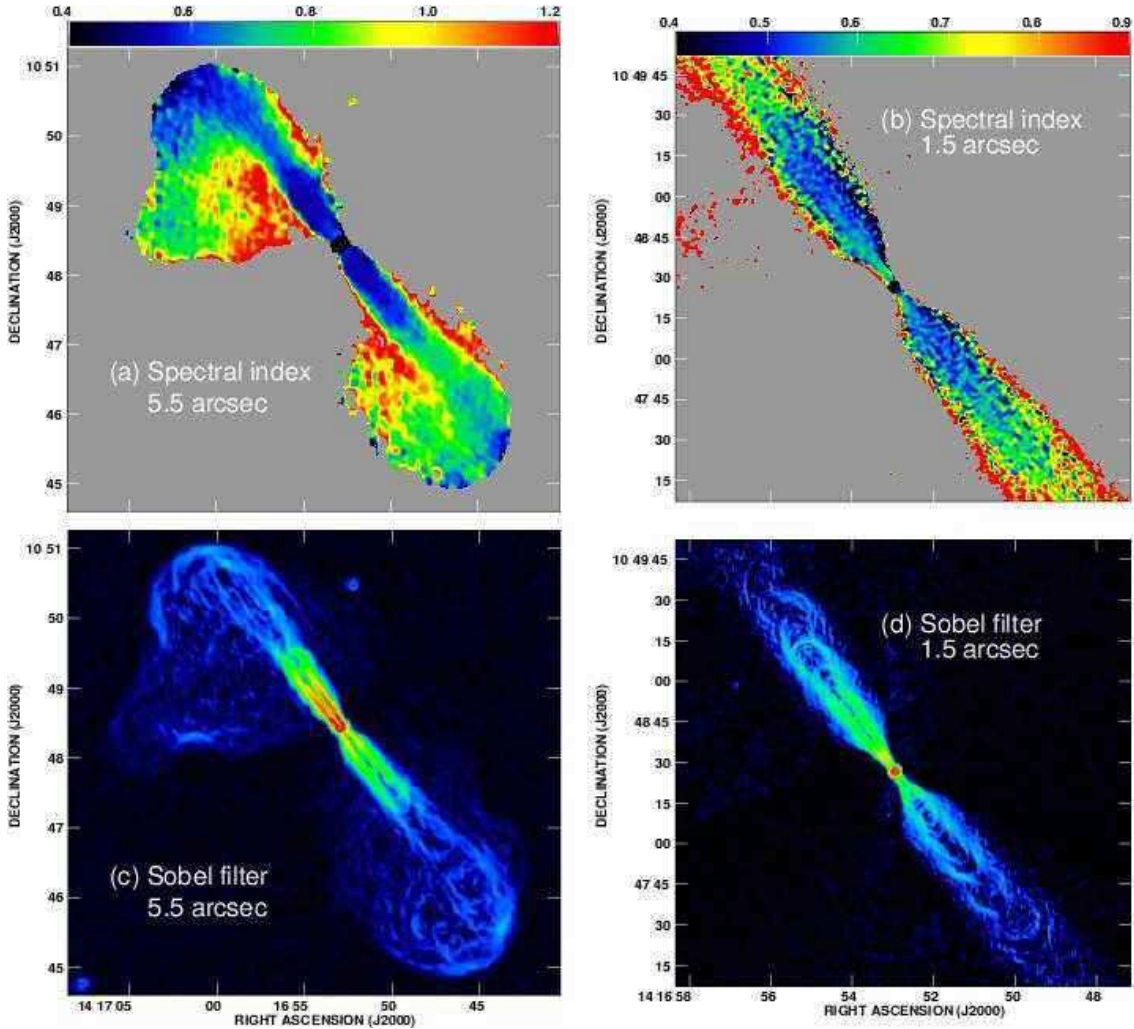}
\caption{False-colour plots of spectral index and intensity gradients. 
(a) Spectral index between 1.48 
and 8.5\,GHz at a resolution of 5.5\,arcsec FWHM. 
The spectral-index range is 0.4--1.2.  
(b) Spectral index between 
1.4 and 8.5\,GHz at a resolution of 1.5\,arcsec FWHM. 
The spectral-index range is 0.4 --
0.9.In panels (a) and (b), values are plotted only if the rms error 
is $<$0.2; blanked pixels are coloured grey.
(c) Sobel-filtered L-band image at 5.5\,arcsec FWHM;
(d) Sobel-filtered 8.5-GHz image at 1.5\,arcsec FWHM.
\label{fig:spec}}
\end{figure*}

The D-configuration observations at 8.5\,GHz nominally sample a largest spatial
scale of only 180\,arcsec. Although the total angular extent of 3C\,296 is
440\,arcsec, the accurate calibration of our dataset and the fact that we have
recovered the total flux of the source imply that we have enough information to
construct a low-resolution image of spectral index, $\alpha$, between 8.5 and
1.48\,GHz. This is shown in Fig.~\ref{fig:spec}(a).  We have also made a
spectral-index image at 1.5 arcsec resolution between 8.5 and 1.41\,GHz; here
the values in the diffuse emission are not reliable, and we show only the inner
jets (Fig.~\ref{fig:spec}b). Sobel-filtered $I$ images at the two resolutions
are shown for comparison in Figs~\ref{fig:spec}(c) and (d).

\begin{figure}
\includegraphics[width=8.5cm]{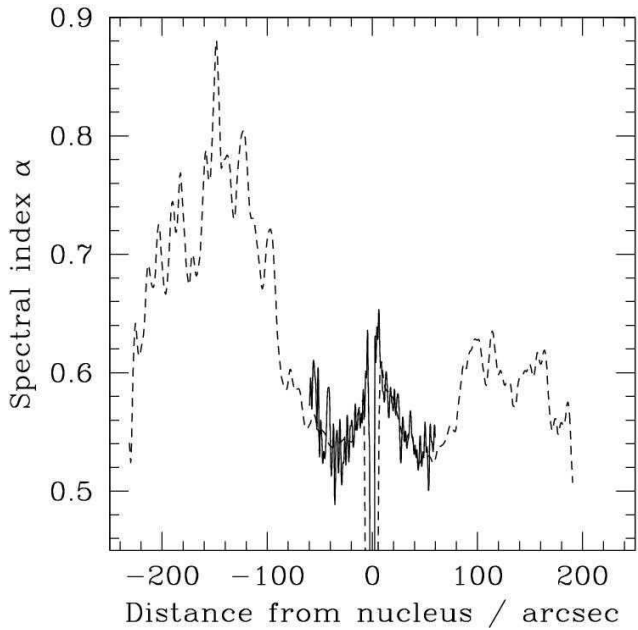}
\caption{Profiles of the spectral index along the jets of 3C\,296 in PA
$-35.^\circ 86$. Dashed line: $\alpha$ measured between 1.48 and 8.5\,GHz at a
resolution of 5.5\,arcsec FWHM, along the jet ridge line from
Fig.~\ref{fig:spec}(a).  Full line: $\alpha$ between 1.41 and 8.5\,GHz at a
resolution of 1.5\,arcsec, averaged over $\pm$1.5\arcsec on either side of the
jet axis, from Fig.~\ref{fig:spec}(b).
\label{fig:alphaprof}}
\end{figure}

Fig.~\ref{fig:spec}(a) shows that the low-brightness regions of the lobes have
significantly steeper spectra than the jets.  Spectral indices in the jets at
5.5-arcsec resolution range from $\alpha = 0.53$ close to the nucleus in both
jets to $\alpha \approx 0.6$ in the main jet and $\alpha \approx 0.8$ in the
counter-jet. The steepest-spectrum diffuse emission visible in
Fig.~\ref{fig:spec}(a) has $\alpha \approx 1.3$. As in other FR\,I objects
\citep{K-SR,KSetal}, the jets appear to be superposed on steeper spectrum lobe
emission, so there must be blending of spectral components.  A steeper-spectrum
component is present on both sides of the main and counter-jets, although it is
more prominent to the SE. It is clearly visible and resolved from
$\approx$17\,arcsec outwards along the main jet. The apparent narrowness of the
steep-spectrum rim within $\approx$50\,arcsec of the nucleus in the counter-jet,
is misleading, however. Much of the lobe emission detected at 1.5\,GHz in this
region (Fig.~\ref{fig:ifull}a) is too faint at 8.5\,GHz to determine a reliable
spectral index and the image in Fig.~\ref{fig:spec}(a) is therefore blanked.
Conversely, any superposed diffuse emission will have a small effect on the
spectral index where the jet is bright. Only within roughly a beamwidth of the
observed edge of the jet are the two contributions comparable, with a total
brightness high enough for $\alpha$ to be determined accurately. The spectrum
therefore appears steeper there (Fig.~\ref{fig:spec}a). With higher sensitivity,
more extended steep-spectrum emission should be detectable around the inner
counter-jet.

\begin{figure}
\includegraphics[width=7.2cm]{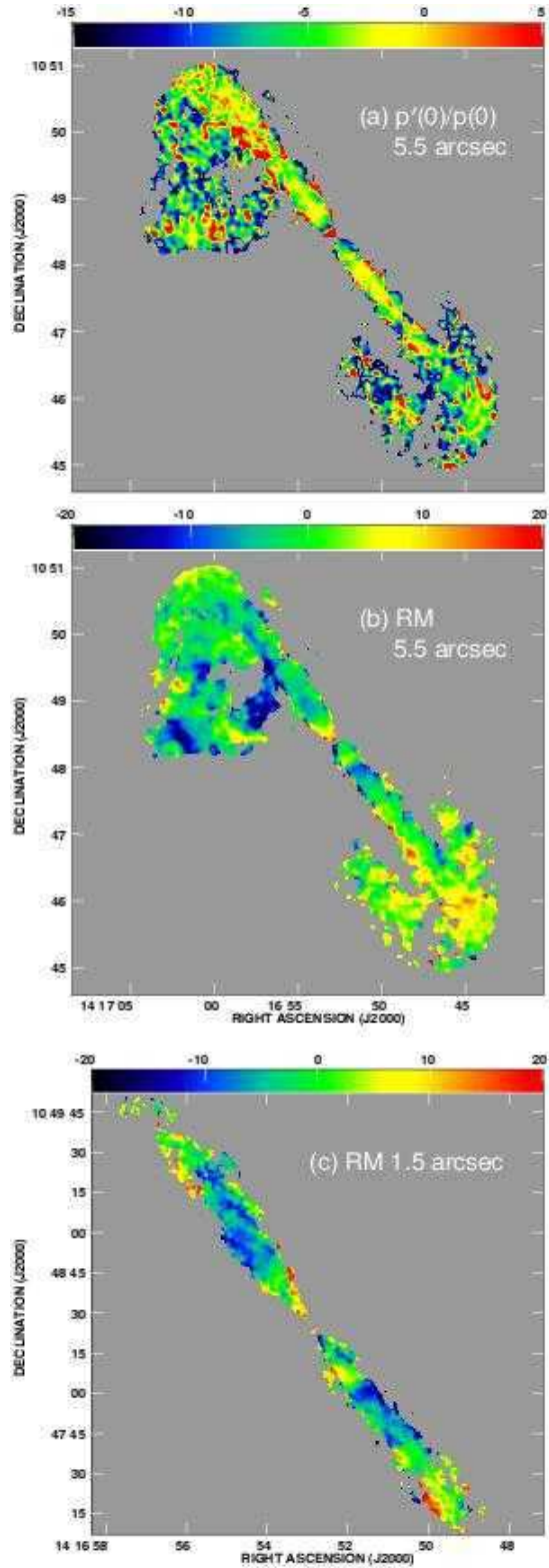}
\caption{(a) Normalized polarization gradient $p^\prime(0)/p(0)$ from a fit to
  $p$ at frequencies of 8.5, 1.45 and 1.50\,GHz over the range $-20$ to
  $+5$\,m$^{-2}$. The resolution is 5.5\,arcsec FWHM. (b) RM image at 5.5-arcsec
  FWHM resolution, from a fit to position angles at the frequencies used in
  panel (a). (c) RM at 1.5\,arcsec from a fit to position angles at 8.5, 1.46
  and 1.39\,GHz.  The labelled wedges for panels (b) and (c) are in units of
  rad\,m$^{-2}$. The areas covered by the panels correspond to
  Figs~\ref{fig:ifull} for (a) and (b) and Fig.~\ref{fig:i1.5} for (c). Blanked
  pixels are coloured grey.
\label{fig:rm}}
\end{figure}

\begin{figure}
\includegraphics[width=8.5cm]{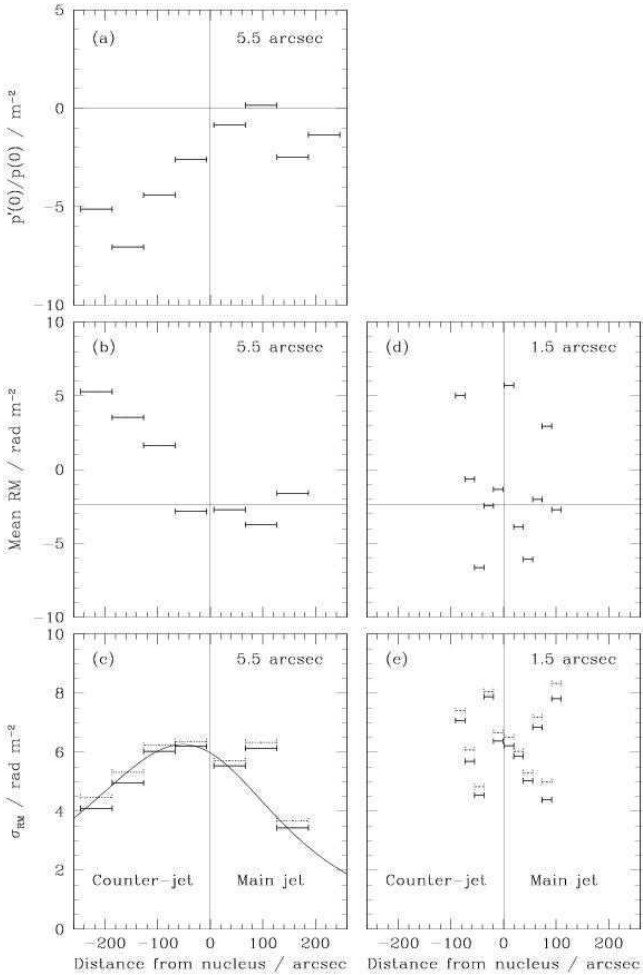}
\caption{Profiles of RM and depolarization along the jet axis, taken to be in PA
  $-35^\circ .86$. Positive and negative distances refer to the main and
  counter-jet sides, respectively. Panels (a) -- (c) show averages for boxes of
  length 60\,arcsec (along the jet axis) and width 39\,arcsec at a resolution of
  5.5\,arcsec FWHM. A $\pm$7.5-arcsec region around the core is excluded. (a)
  Mean normalized polarization gradient $\langle p^\prime(0)/p(0) \rangle$. (b)
  Mean RM. The horizontal line shows the estimated Galactic RM. (c) Rms RM. The
  full lines show $\sigma_{\rm RM}$ after a first-order correction for fitting
  error, the dotted lines show the uncorrected values $\sigma_{\rm RM raw}$ and
  the curve shows a simple model fitted by eye to the data (see text). (d) and
  (e) As (b) and (c), but at a resolution of 1.5\,arcsec FWHM. The boxes are
  18\,arcsec long and are extended transverse to the jet to include all of the
  unblanked emission in Fig.~\ref{fig:rm}(c). The excluded region around the
  core is $\pm 1.5$\,arcsec.
\label{fig:rmprofs}}
\end{figure}

Flatter-spectrum emission can be traced beyond the sharp outer bends in both
jets, suggesting that the flow in the outer part of the radio structure remains
coherent even after it deflects through a large angle, at least in
projection.\footnote{Note, however, that the very flat spectral indices $\alpha \la
  0.5$ seen at both ends of the source may be unreliable because
  of the large correction for primary beam attenuation required at 8.5\,GHz.}
A comparison between the spectral-index image and a Sobel-filtered L-band image
(Figs~\ref{fig:spec}a and c) shows that the flatter-spectrum regions are bounded by
high brightness gradients; they can therefore be separated both morphologically and
spectrally from the surrounding steep-spectrum emission.  In the NE lobe, the
flatter-spectrum ``extension'' of the jet can be traced until it terminates at a
region with an enhanced intensity (and intensity gradient) on the South edge of
the lobe (Figs~\ref{fig:ifull}a, \ref{fig:spec}a and c).  

The spectral structure therefore suggests that the jets deflect when they reach
the ends of the lobes rather than disrupting completely. In FR\,I sources such
as 3C\,296, no strong shocks (hot-spots) are formed, implying that the
generalised internal Mach number of the flow, ${\cal M} =
\Gamma\beta/\Gamma_{\rm s}\beta_{\rm s} \la 1$, where $\beta_{\rm s}c$ is the
internal sound speed and $\Gamma_{\rm s} = (1-\beta_{\rm s}^2)^{-1/2}$
\citep{Kon80}. For an ultrarelativistic plasma, $\beta_{\rm s} = 3^{-1/2}$, so
strong shocks need not occur at the bends even if the flow remains mildly
relativistic (e.g.\ we infer a transonic flow decelerating from $\beta \approx
0.8$ to $\beta \approx 0.2$ in 3C\,31; \citealt{LB02b}).  In 3C\,296, our models
imply that $\beta \approx 0.4$ on-axis at a projected distance of 40\,arcsec
from the nucleus, with a possible deceleration to $\beta \la 0.1$ at larger
distances (Section~\ref{results:vel}).

It is plausible that the spectra in the lobe regions of 3C\,296 have been
steepened by a combination of synchrotron, inverse Compton and adiabatic losses
but it is not clear from the spectral-index image whether the jets and their
extensions propagate within the steep-spectrum lobes (as in the standard model
for FR\,II sources) or appear superimposed on them.

At higher resolution (Fig.~\ref{fig:spec}b), there are variations in $\alpha$
both along and transverse to the jets. Fig.~\ref{fig:alphaprof} shows a
longitudinal profile of spectral index along the axis within 60\,arcsec of the
nucleus, where the superposed diffuse, steep-spectrum emission has a negligible
effect. There is good agreement between the spectral indices measured at 1.5 and
5.5-arcsec resolution in this area.  The spectral index close to the nucleus in
both jets is $\approx$0.62. The spectrum then flattens with distance from the
nucleus to $\alpha \approx 0.53$ at 40\,arcsec. Further from the nucleus, the
spectrum steepens again (Fig.~\ref{fig:spec}b), but lobe contamination becomes
increasingly important and it is not clear that we are measuring the true
spectrum of the jet. The flattening with increasing distance in the first
40\,arcsec of both jets is in the opposite sense to that expected from
contamination by lobe emission, however.  The few other FR\,I jets observed with
adequate resolution and sensitivity (none of which are superposed on lobes) show
a similar spectral flattening with distance: PKS\,1333$-$33 \citep{KBE}, 3C\,449
\citep{K-SR}, NGC\,315 \citep{ngc315ls}, and NGC\,326 (Murgia, private
communication).

A comparison of Figs~\ref{fig:spec}(b) and (d) shows that the spectral index is
flatter in both jets roughly where the more distinct arcs are seen, but the arcs
are not recognisable individually on the spectral index images.  They contain
only a few tens of percent of the total jet emission along their lines of sight,
so a small difference in spectral index between them and their surroundings
would not be detectable.  The apparent narrowness of the steeper-spectrum
($\alpha \approx 0.9$) rim observed at the edges of the jets in
Fig.~\ref{fig:spec}(b) results from the superposition of jet and lobe emission,
just as at lower resolution.  This effect confuses any analysis of intrinsic
transverse spectral gradients in the jets, such as those in the flaring region
of NGC\,315 \citep{ngc315ls}, but there are hints that the intrinsic spectrum
flattens slightly away from the axis within $\approx$30\,arcsec of the nucleus
in 3C\,296 (Fig.~\ref{fig:spec}b).

At 0.4-arcsec resolution, we measure $\alpha \approx 0.67$ between 4.9 and
8.5\,GHz for the bright region of the main jet between 2 and 8 arcsec from the
nucleus and $\alpha \approx 0.57$ for the bright knot at the base of the
counter-jet.

\begin{figure*}
\includegraphics[width=15cm,angle=-90]{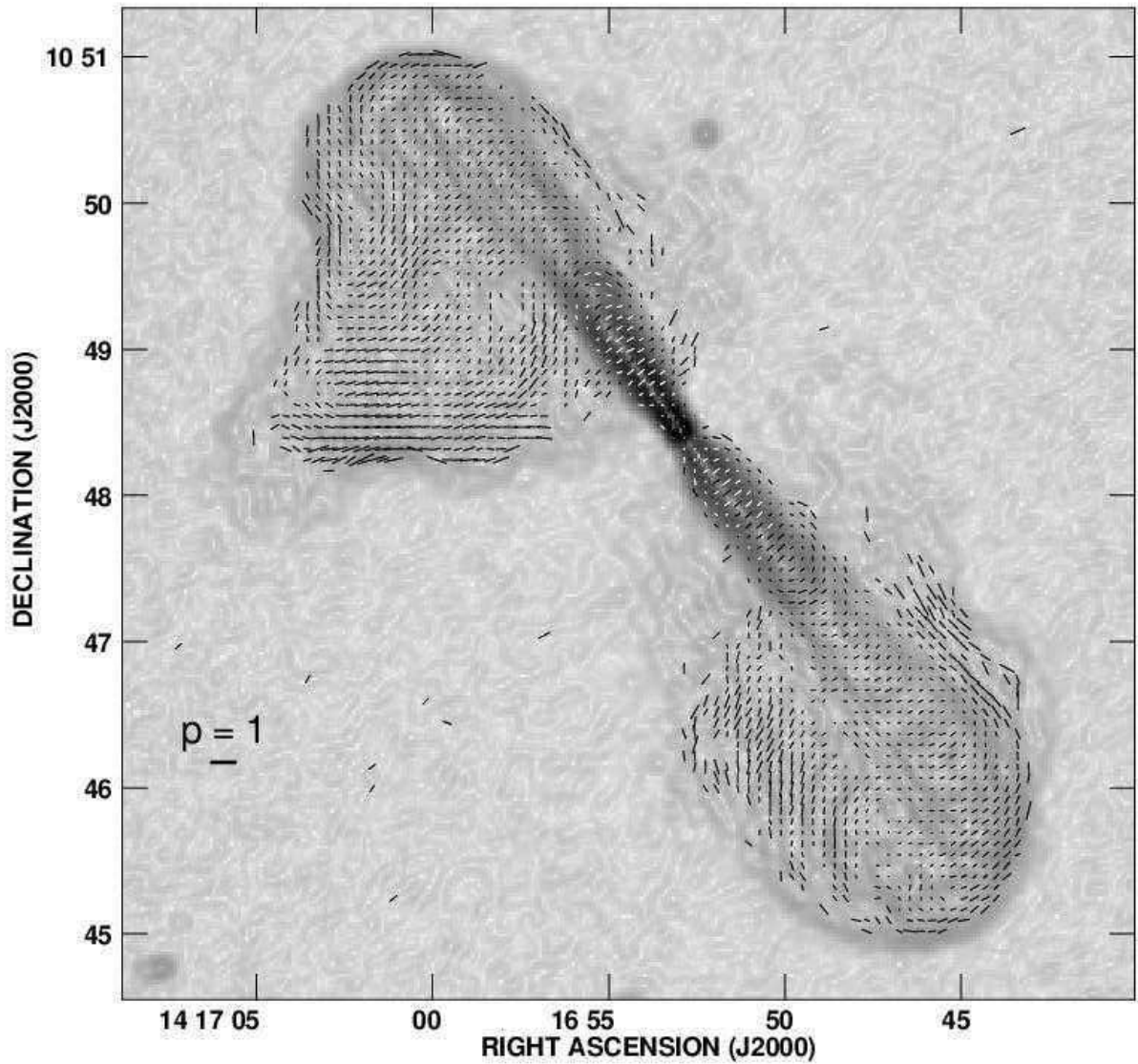}
\caption{Vectors with lengths proportional to the degree of polarization at
  8.5\,GHz and directions along the apparent magnetic field, superimposed on a
  grey-scale of the total intensity gradient (Sobel-filtered image) at 1.5\,GHz.
  The resolution is 5.5\,arcsec FWHM.  The vector directions are derived from
  the 3-frequency RM fits where there is adequate signal-to-noise ratio;
  elsewhere, the 8.5-GHz position angles are corrected using the mean RM over
  the area of the image. Vectors are plotted where $P \geq 2\sigma_P$ and $I
  \geq 5 \sigma_I$.
\label{fig:polvecslo}}
\end{figure*}

\subsection{Faraday rotation and depolarization}
\label{RM}

Comparison of lower-resolution images at 1.7 and 0.6\,GHz showed that the SW
(counter-jet) lobe depolarizes more rapidly than the NE (main jet) lobe
\citep{GHS}. We derived the polarization gradient $p^\prime(0) =
dp/d(\lambda^2)$ and the polarization at zero wavelength, $p(0)$ from a linear
fit to the degree of polarization as a function of $\lambda^2$, $p(\lambda^2)
\approx p(0) + p^\prime(0)\lambda^2$, using 5.5-arcsec resolution images at 8.5,
1.50 and 1.45\,GHz. The linear approximation is adequate for the low
depolarization seen in 3C\,296 and allows us to use images at more than two
frequencies to reduce random errors. We show an image of $p^\prime(0)/p(0)$
(closely related to depolarization) in Fig.~\ref{fig:rm}(a). There is a gradient
along the jets, in the sense that the outer counter-jet is more depolarized than
the main jet; we show a profile along the axis in
Fig.~\ref{fig:rmprofs}(a). Both lobes show significant depolarization: $\langle
p^\prime(0)/p(0)\rangle = -3.7$\,m$^{-2}$ and $-5.8$\,m$^{-2}$ for the NE and SW
lobes, respectively, corresponding to depolarizations of 0.84 and 0.75 at
1.45\,GHz. As can be seen from Figs~\ref{fig:rm}(a) and \ref{fig:rmprofs}(a),
the diffuse emission in the NE lobe is significantly more depolarized ($\langle
p^\prime(0)/p(0)\rangle = -4.8$\,m$^{-2}$) than the main jet. In the SW lobe,
the depolarization of the diffuse emission ($\langle p^\prime(0)/p(0)\rangle =
-6.2$\,m$^{-2}$) is comparable to that of the counter-jet measured further than
100\,arcsec from the nucleus (Fig.~\ref{fig:rmprofs}a).

The integrated Faraday rotation measure (RM) for 3C\,296 is $-3 \pm
2$\,rad\,m$^{-2}$ \citep{SKB} and a mean value of 0\,rad\,m$^{-2}$ with an rms
dispersion of 4\,rad\,m$^{-2}$ was derived by \citet{LPR86} from three-frequency
imaging at a resolution of $6 \times 22$\,arcsec$^2$ FWHM.  We imaged the
distribution of RM at 5.5 arcsec resolution, fitting to ${\bf E}$-vector
position angles, $\chi$,  at 8.5, 1.50 and 1.45\,GHz.  Given the closeness of the two
lower frequencies, we have poor constraints on the linearity of the $\chi$ --
$\lambda^2$ relation, but the values of RM are sufficiently small that we can be
sure that there are no $n\pi$ ambiguities (this is confirmed by comparison with
the 2.7-GHz images of \citealt{BLP}).  The resulting RM image is shown in
Fig.~\ref{fig:rm}(b). The mean RM is $-0.4$\,rad\,m$^{-2}$, with an rms of
6.1\,rad\,m$^{-2}$ and a total spread of $\pm$20\,rad\,m$^{-2}$. Profiles of
mean and rms RM along the jet axis are shown in Figs~\ref{fig:rmprofs}(b) and
(c). We made a first-order correction to the rms RM, $\sigma_{\rm RM raw}$, by
subtracting the fitting error $\sigma_{\rm fit}$ in quadrature to give
$\sigma_{\rm RM} = (\sigma_{\rm RM raw}^2 - \sigma_{\rm fit}^2)^{1/2}$. Both raw
and corrected values are plotted in Fig.~\ref{fig:rmprofs}(c).  There is a clear
large-scale gradient in RM across the counter-jet lobe, roughly aligned with the
axis and with an amplitude of $\approx$8\,rad\,m$^{-2}$. Fluctuations on a range
of smaller scales are also evident. The rms RM peaks close to the nucleus and
decreases by a factor of 1.5 -- 2 by 200\,arcsec away from it
(Fig.~\ref{fig:rmprofs}c).

\begin{figure}
\includegraphics[width=8.5cm]{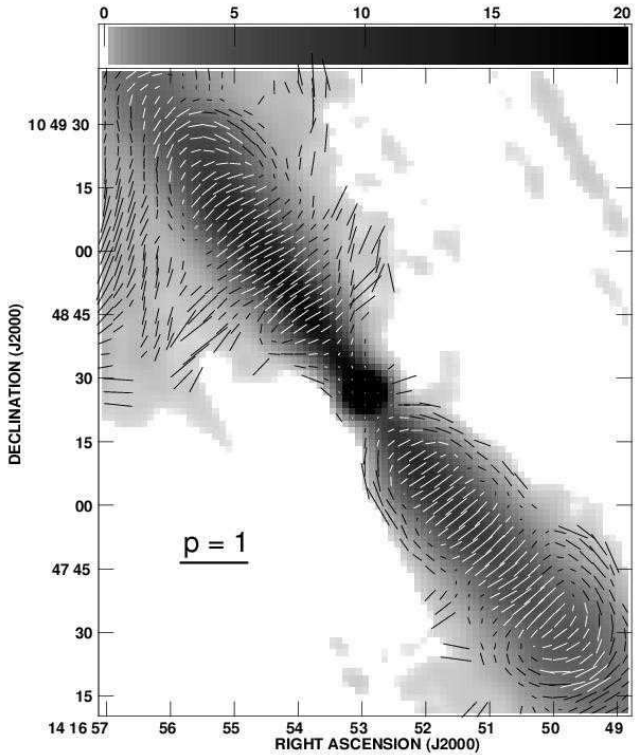}
\caption{Vectors with lengths proportional to the degree of polarization at
  8.5\,GHz and directions along the apparent magnetic field, superimposed on
  a grey-scale of total intensity at the same frequency. A square root transfer
  function has been used to emphasise the diffuse emission around the main
  jet. The resolution is 5.5 arcsec FWHM. The vector directions are derived from
  3-frequency RM fits where there is adequate signal-to-noise ratio; elsewhere,
  the 8.5-GHz position angles are corrected using the mean RM over the area of
  the image. Vectors are plotted where $P \geq 2\sigma_P$ and $I \geq 5
  \sigma_I$.
\label{fig:ivecdeep}}
\end{figure}

At 1.5-arcsec resolution, we made 3-frequency RM fits to images at 1.39,
1.46 and 8.5\,GHz (Fig.~\ref{fig:rm}c)\footnote{A position-angle image
at 1.41\,GHz (derived using only the A-configuration data at this frequency)
is consistent with this RM fit but too noisy to improve it}. Note that the largest scale
sampled completely in this image is $\approx$40\,arcsec.  Profiles of mean
and rms RM at this resolution are shown in Figs~\ref{fig:rmprofs}(d) and
(e). There are no systematic trends in either quantity within 100\,arcsec of the
nucleus. 

Our data are qualitatively consistent with the results of \citet{GHS} in the
sense that we see somewhat more depolarization in the counter-jet lobe at
5.5-arcsec resolution. The corresponding profile of RM fluctuations is quite
symmetrical, however. If we calculate the depolarization expected with the beam
used by \citet{GHS} between 1.7 and 0.6\,GHz using the observed degree of
polarization at 1.4\,GHz and the RM image at 5.5-arcsec resolution, we find no
gross differences between the lobes. This analysis neglects spectral
variations across the source and depolarization due to any mechanism other than
foreground Faraday rotation on scales $\ga$5.5\,arcsec, however.
 
A simple model in which the variations in RM result from foreground field
fluctuations in a spherically-symmetric hot gas halo is reasonable for 3C\,296:
the model profile plotted in Fig.~\ref{fig:rmprofs}(c) was derived following the
prescription in \citet{ngc315ls}, assuming an angle to the line of sight of
$58^\circ$ (Section~\ref{results:geom}) and fitted by eye to the data. For the
derived core radius $r_c = 300$\,arcsec, the asymmetries in the predicted
$\sigma_{\rm RM}$ profile for angles to the line of sight $\ga 50^\circ$ are
small and reasonably consistent with our observations.  A full analysis must,
however, include the gradient of RM across the counter-jet lobe and the enhanced
depolarization seen in that region, neither of which follow the same profile as
the rms RM. The large-scale gradient is extremely unlikely to be Galactic, as
the total Galactic RM estimated from the models of \citet{DC05} is only
$-2.4$\,rad\,m$^{-2}$ and fluctuations on 200-arcsec scales are likely to be
even smaller. The gradient must, therefore, be local to 3C\,296 (a similar
large-scale RM gradient, albeit of much larger amplitude, was observed in
Hydra~A by \citealt{HydraA}). We defer a quantitative analysis of the Faraday
rotation and depolarization until XMM-Newton imaging of the hot gas on the scale
of the lobes is available for 3C\,296.

\subsection{Apparent magnetic field}

\begin{figure*}
\includegraphics[width=17cm]{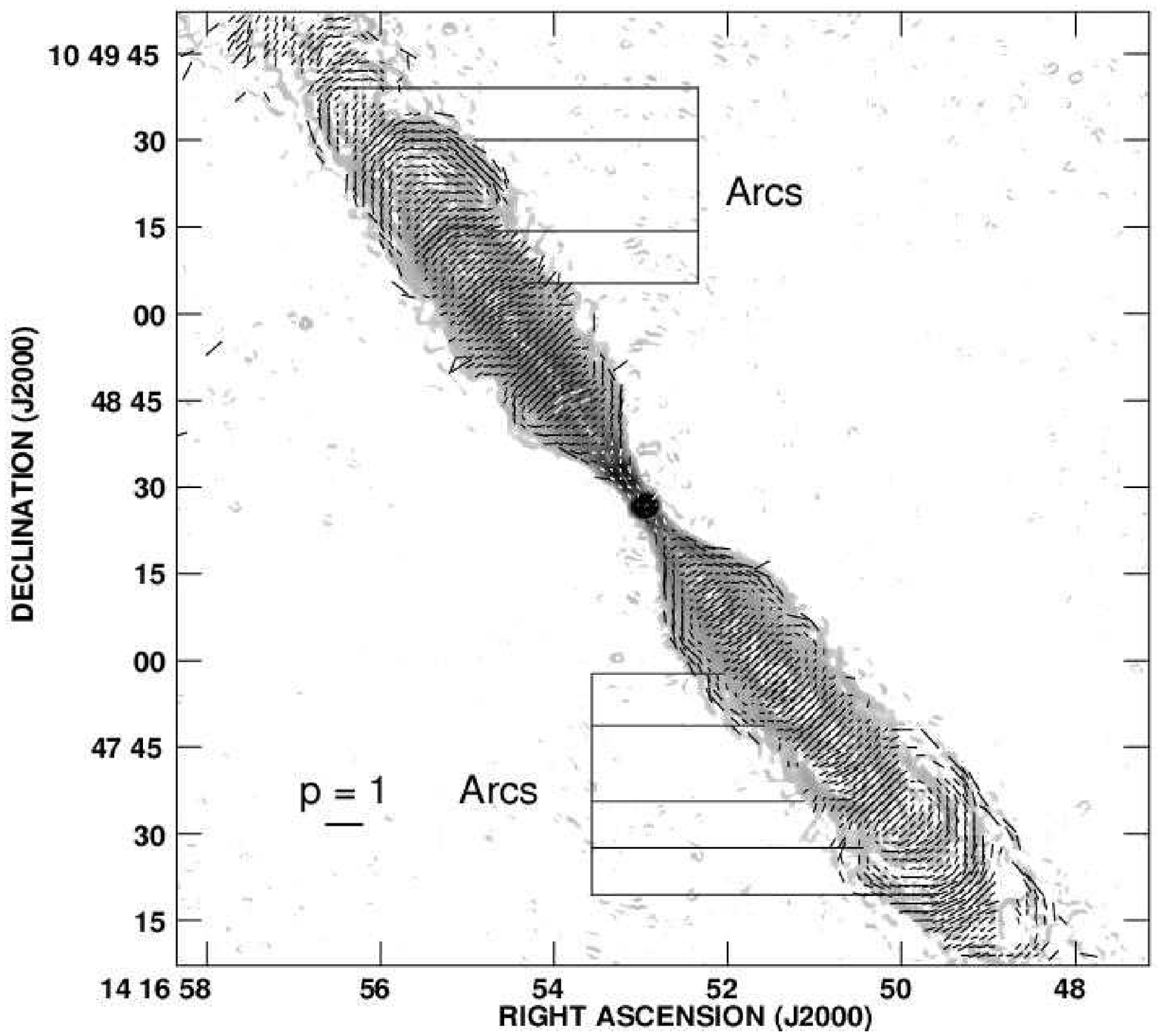}
\caption{Vectors with lengths proportional to the degree of polarization at
  8.5\,GHz and directions along the apparent magnetic field, superimposed on a
  grey-scale of Sobel-filtered total intensity at the same frequency. The resolution is 1.5
  arcsec FWHM. The vector directions are derived from 3-frequency RM fits where
  there is adequate signal-to-noise ratio; elsewhere, the 8.5-GHz position
  angles are corrected using the mean RM of the image. Vectors are plotted where
  $P \geq 3\sigma_P$ and $I \geq 5 \sigma_I$. The ``arcs'' are those marked in
  Fig.~\ref{fig:i1.5}(b). 
\label{fig:ivec1.5}}
\end{figure*}

We show the direction of the {\sl apparent magnetic field}, rotated from the
zero-wavelength {\bf E}-vector position angles by 90$^\circ$. Over much of the
observed region, this can be derived directly from the $\chi$ -- $\lambda^2$
fit.  The 8.5-GHz images generally have more points with significant
polarization than those at lower frequencies, however. At points with
significant polarized signal at 8.5\,GHz but no RM measurement, we determined
the apparent magnetic-field direction using the observed 8.5-GHz position angle
and the mean RM for the region. The maximum error introduced by this procedure
is $\approx 1.4^\circ$ provided that the RM distribution is no wider than in
Fig.~\ref{fig:rm}(b).  The apparent field directions are shown for the whole
source at 5.5\,arcsec FWHM resolution in Fig.~\ref{fig:polvecslo}, in more
detail for a small region around the nucleus at the same resolution in
Fig.~\ref{fig:ivecdeep}, and for the jets at 1.5-arcsec FWHM resolution in
Fig.~\ref{fig:ivec1.5}.  The vector lengths in these plots are proportional to
the degree of polarization at 8.5\,GHz, corrected to first order for Ricean bias
\citep{WK}; this is equal to $p(0)$ within the errors at all points where we
have sufficient signal-to-noise to estimate depolarization (Section~\ref{RM}).

As noted by \citet{LP91}, the apparent magnetic field in the lobe emission is
circumferential, with a high degree of polarization at the edges of the lobes.
The magnetic field directions are well aligned with the ridges of the steepest
intensity gradients, as shown by the superposition on the Sobel-filtered L-band
image (Fig.~\ref{fig:polvecslo}).  The jets show mainly transverse apparent
field at 5.5\,arcsec resolution, but the low-brightness regions around them do
not all exhibit the same pattern of polarization. Those to the NW of both jets
show $p \ga 0.4$ with the field ordered parallel to the jet axis. Those
to the SE of the main jet are comparably polarized but with the magnetic field
oblique to the jet axis while those to the SE of the counter-jet are only weakly
polarized.  Both lobes contain polarized structure related to the spectral-index
structure evident in Fig.~\ref{fig:spec}(a).  The field structure in the S lobe
is suggestive of a flow that turns through 180$^\circ$ but retains a
transverse-field configuration after the bend.  That in the N lobe is suggestive
of a jet extension whose magnetic configuration changes from parallel
immediately after the bend to transverse.  The emission at the S edge of the
lobe where there is a locally strong intensity gradient also exhibits a high
($p \ga 0.6$) linear polarization with the magnetic field tangential to the lobe
boundary, i.e. transverse to the jet ``extension'' suggested by our spectral
data.

At 1.5-arcsec resolution, the jet field structure is revealed in more detail
(Fig.~\ref{fig:ivec1.5}). Longitudinal apparent field is visible at the edges of
both jets, most clearly close to the nucleus and at $\approx$60\,arcsec from it.
There are enhancements in polarization associated with the two outermost arcs in
each jet: in all four cases the apparent field direction is perpendicular to the
maximum brightness gradient.

The 8.5-GHz $Q$ and $U$ images at 0.75 and 0.25 arcsec FWHM used for modelling
have been corrected for Faraday rotation using the 1.5-arcsec RM images
interpolated onto a finer grid. The corresponding apparent field directions are
discussed in Section~\ref{Pcomp}.

\section{The model}
\label{model}

\subsection{Assumptions}

Our principal assumptions are as as in LB, CL and CLBC:
\begin{enumerate}
\item FR\,I jets may be modelled as intrinsically symmetrical, antiparallel,
  axisymmetric, stationary laminar flows. Real flows are, of course, much more
  complex, but our technique will still work provided that the two jets are
  statistically identical over our averaging volumes.
\item The jets contain relativistic particles with an energy spectrum $n(E)dE =
n_0 E^{-(2\alpha+1)}dE$ and an isotropic pitch-angle distribution, emitting
optically-thin synchrotron radiation with a frequency spectral index $\alpha$.
We adopt a spectral index of $\alpha = 0.60$ derived from an average of the
spectral-index image in Fig.~\ref{fig:spec}(b) over the modelled region. The
maximum degree of polarization is then $p_0 = (3\alpha+3)/(3\alpha+5) = 0.705$.
\item The magnetic field is tangled on small scales, but anisotropic
  \citep{Laing81,BBR,LB02a,LCB}. 
\end{enumerate}

\subsection{Outline of the method}
\label{outline-method}

It is well known that synchrotron radiation from intrinsically symmetrical,
oppositely directed bulk-relativistic outflows will appear one-sided as a result
of Doppler beaming. The jet/counter-jet ratio, $I_{\rm j}/I_{\rm cj} =
[(1+\beta\cos\theta)/(1-\beta\cos\theta)]^{2+\alpha}$ for a constant-speed,
one-dimensional flow with isotropic emission in
the rest frame.  The key to our approach is that aberration also acts
differently on linearly-polarized radiation from the approaching and receding
jets, so their observed polarization images represent two-dimensional
projections of the magnetic-field structure viewed from different directions
$\theta_{\rm j}^\prime$ and $\theta_{\rm cj}^\prime$ in the rest frame of the
flow: $\sin\theta_{\rm j}^\prime = \sin\theta [\Gamma(1-\beta\cos\theta)]^{-1}$
and $\sin\theta_{\rm cj}^\prime = \sin\theta
[\Gamma(1+\beta\cos\theta)]^{-1}$. This is the key to breaking the degeneracy
between $\beta$ and $\theta$ and to estimating the physical parameters of the
jets \citep{CL}.

We have developed a parameterized description of the jet geometry, velocity
field, emissivity and magnetic-field ordering. We calculate the emission from a
model jet in $I$, $Q$ and $U$ by numerical integration, taking full account of
relativistic aberration and anisotropic rest-frame emission, convolve to the
appropriate resolution and compare with the observed VLA images.  We optimize
the model parameters using the downhill simplex method with $\chi^2$ (summed
over $I$, $Q$ and $U$) as a measure of goodness of fit (note that the modelling
is not affected by Ricean bias as we fit to Stokes $I$, $Q$ and $U$
directly). The method is described fully by LB, CL and CLBC.

Our method is restricted to the vicinity of the nucleus, where we expect that
relativistic effects dominate over intrinsic or environmental asymmetries. The
modelled region for 3C\,296 (indicated by the box on Fig.~\ref{fig:ihires}a) is
limited in extent by slight bends in the jets at $\approx$45\,arcsec from the
nucleus. Contamination by surrounding lobe emission also becomes significant at
larger distances.
 
\subsection{Functional forms for geometry, velocity, magnetic field and
  emissivity}

The functional forms used for velocity, emissivity and field ordering are
identical to those described by CLBC except for a very small change to the
emissivity profile (Section~\ref{emiss-parms}). We briefly summarize the
concepts in this Section and give the functions in full for reference in
the Appendix (Table~\ref{tab:param}).

\begin{figure}
\includegraphics[width=8.5cm]{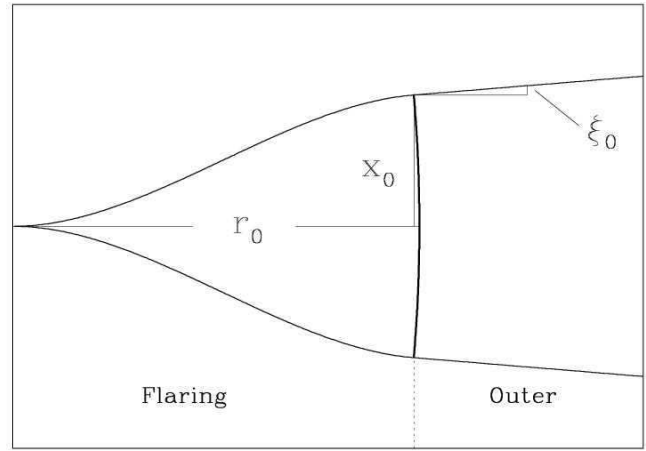}
\caption{The geometry of the jet, showing the flaring and outer regions and
  the quantities which define the shape of its outer surface. 
\label{fig:geomsketch}}
\end{figure}

\subsubsection{Geometry}
\label{geom-parms}

The assumed form for the jet geometry (Fig.~\ref{fig:geomsketch}) is that used
by CL and CLBC. The jet is divided into a {\em flaring region} where it first
expands and then recollimates, and a conical (almost cylindrical) {\em outer
region}. The geometry is defined by the angle to the line of sight, $\theta$,
the distance from the nucleus of the transition between flaring and outer
regions, $r_0$, the half-opening angle of the outer region, $\xi_0$ and the
width of the jet at the transition, $x_0$ (all fitted parameters).

We assume laminar flow along streamlines defined by an index $s$ ($0 \leq s \leq
1$) where $s = 0$ corresponds to the jet axis and $s = 1$ to its edge. In the
outer region, the streamlines are straight and in the flaring region the
distance of a streamline from the axis of the jet is modelled as a cubic
polynomial in distance from the nucleus. Distance along a streamline is measured
by the coordinate $\rho$, which is exactly equal to distance from the nucleus
for the on-axis streamline. The streamline coordinate system is defined in
Appendix~\ref{functional-forms}. 

\subsubsection{Velocity}
\label{vel-parms}

The form of the velocity field is as used by CL and CLBC. The on-axis profile is
divided into three parts: roughly constant, with a high velocity $\beta_1$ close
to the nucleus; a linear decrease and a roughly constant but lower velocity
$\beta_0$ at large distances.  The profile is defined by four free parameters:
the distances of the two boundaries separating the regions and the
characteristic inner and outer velocities. Off-axis, the velocity is calculated
using the same expressions but with truncated Gaussian transverse profiles
falling to fractional velocities varying between $v_1$ and $v_0$
(Table~\ref{tab:param}).

\subsubsection{Magnetic field}
\label{field-parms}

We define the rms components of the magnetic field to be $\langle
B_l^2\rangle^{1/2}$ (longitudinal, parallel to a streamline) and $\langle
B_t^2\rangle^{1/2}$ (toroidal, orthogonal to the streamline in an azimuthal
direction). We initially included a radial field component in the model, exactly
as for NGC\,315 (CLBC). The best-fitting solution had a small radial component
on the jet axis, but not elsewhere, and the fitted coefficients were everywhere
consistent with zero. The improvement in $\chi^2$ resulting from inclusion of a
radial component was also very small (see Section~\ref{fit-details}). We
therefore present only models in which the radial field component is zero
everywhere.  The longitudinal/toroidal field ratio varies linearly with
streamline index as in CLBC (Table~\ref{tab:param}) .

\begin{figure*}
\includegraphics[width=17cm]{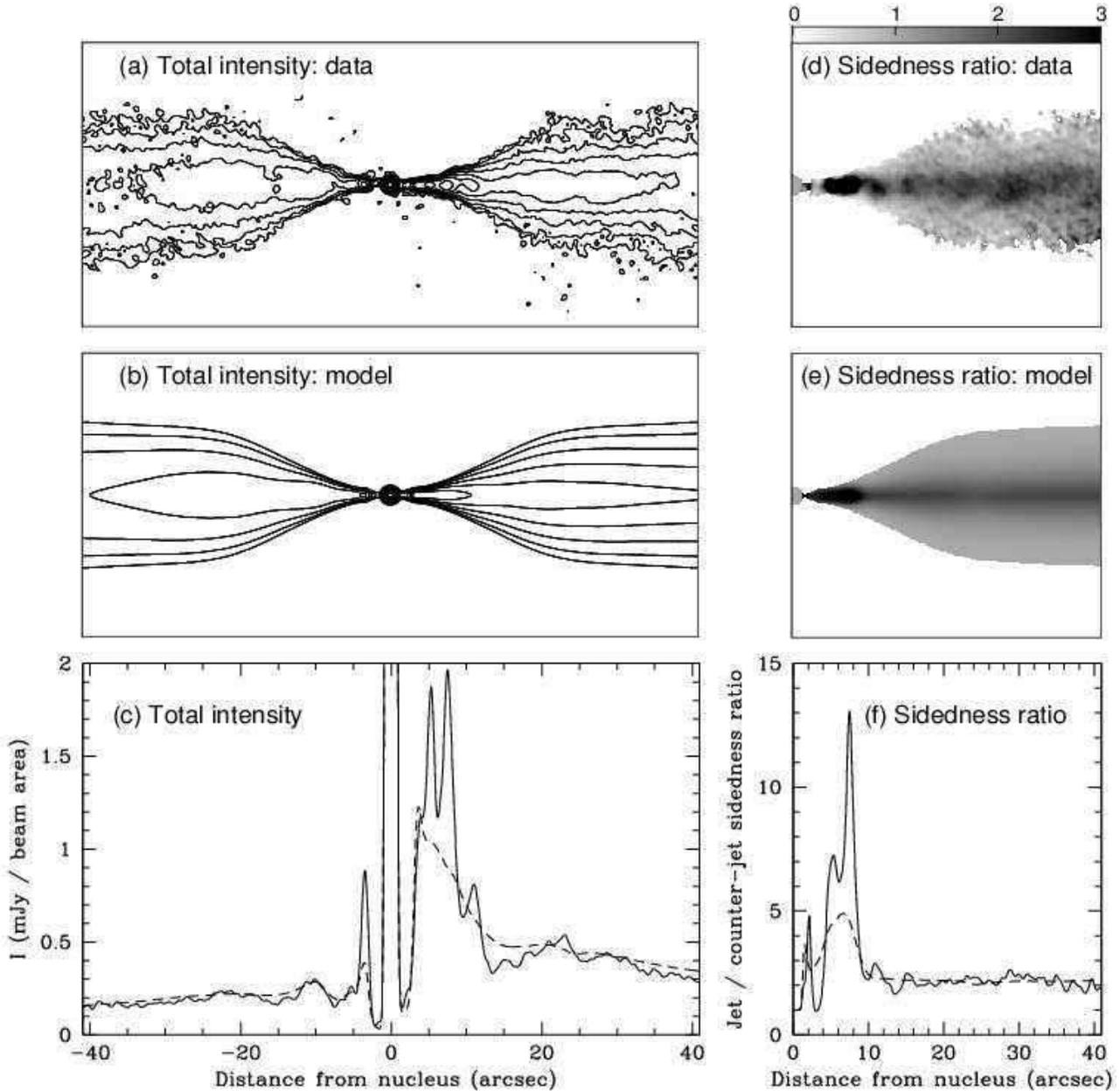}
\caption{Comparison between data and model at a resolution of 0.75\,arcsec. The
  displayed region is $\pm$40.95\,arcsec from the nucleus along the jet axis and
  the scale is indicated at the bottom of the Figure. (a) Observed contours. The
  levels are $-1$, 1, 2, 4, 8, 16, 32, 64, 128, 256, 512, 1024, 2048, 4096
  $\times$ 20\,$\mu$Jy\,beam$^{-1}$. (b) Model contours, with the same levels as
  in panel (a). (c) Profiles of total intensity along the jet axis. Full line:
  data; dashed line: model. (d) Grey-scale of observed jet/counter-jet sidedness
  ratio, in the range 0 -- 3, chosen to show the variations in the outer parts
  of the modelled region. (d) Grey-scale of model sidedness ratio. (e) Profiles
  of sidedness ratio along the jet axis. Full line: data; dashed line: model.  
\label{fig:icomplo}}
\end{figure*}

\subsubsection{Emissivity}
\label{emiss-parms}

We take the proper emissivity to be $\epsilon h$, where $\epsilon$ is the
emissivity in $I$ for a magnetic field $B = \langle B_l^2 +
B_t^2\rangle^{1/2}$ perpendicular to the line of sight and $h$ depends on field
geometry: for $I$, $0 \leq h \leq 1$ and for $Q$ and $U$ $-p_0 \leq h \leq
+p_0$.  We refer to $\epsilon$, loosely, as `the emissivity'. For a given
spectral index, it is a function only of the rms total magnetic field and the
normalizing constant of the particle energy distribution, $\epsilon \propto n_0
B^{1+\alpha}$.

As in CL and CLBC, the majority of the longitudinal emissivity profile is
modelled by three power laws $\epsilon \propto \rho^{-E}$, with indices $E =
E_{\rm in}$, $E_{\rm mid}$ and $E_{\rm out}$ (in order of increasing distance
from the nucleus).  The inner emissivity region corresponds to the faint inner
jets before what we will call the {\em brightening point}.  The middle region
describes the bright jet base and the outer region the remainder of the modelled
area.  We have adopted slightly different ad hoc functional forms for the short
connecting sections between these regions in CL, CLBC and the present paper. For
3C\,296, we allow a discontinuous increase in emissivity at the end of the
innermost region, followed by a short section with power-law index $E_{\rm
knot}$, introduced to fit the first knots in the main and counter-jets. We adopt
truncated Gaussian profiles for the variation of emissivity with streamline
index. The full functional forms for the emissivity (with the names of some
variables changed from CLBC in the interest of clarity) are again given in
Table~\ref{tab:param}.

\subsection{Fitting, model parameters and errors}
\label{fit-details}

The noise level for the calculation of $\chi^2$ for $I$ is estimated as
$1/\sqrt{2}$ times the rms of the difference between the image and a copy of
itself reflected across the jet axis. For linear polarization, the same level is
used for both $Q$ and $U$. This is the mean of $1/\sqrt{2}$ times the rms of the
difference image for $Q$ and the summed image for $U$, since the latter is
antisymmetric under reflection for an axisymmetric model. This prescription is
identical to that used by LB, CL and CLBC. The region immediately around the core is
excluded from the fit.

We fit to the 0.25-arcsec FWHM images within 8.75\,arcsec of the nucleus and to
the 0.75-arcsec FWHM images further out.  In doing so, we implicitly assume that
the jet structure visible at 8.5\,GHz is not significantly contaminated by
superposed lobe emission. For total intensity, this is justified by the
faintness of the lobe emission close to the jet even at much lower resolution
(Fig.~\ref{fig:ivecdeep}); we discuss the possibility of contamination of the
polarized jet emission in Section~\ref{badfit}.  In our initial optimizations,
the quantities defining the shape of the jet projected on the plane of the sky
were allowed to vary. These were then fixed and the remaining parameters were
optimized, using only data from within the outer isophote of the model.

We derive rough uncertainties, as in LB, CL and CLBC, by varying individual
parameters until the increase in $\chi^2$ corresponds to the formal 99\%
confidence level for independent Gaussian errors. These estimates are crude
(they neglect coupling between parameters), but in practice give a good
representation of the range of qualitatively reasonable models.  The number of
independent points (2588 in each of 3 Stokes parameters) is sufficiently large
that we are confident in the main features of the model. The reduced
$\chi^2_{\rm red} = 0.72$, suggesting that our noise model is oversimplified.  A
model with a radial field component and a more general variation of
field-component ratios with streamline index has $\chi^2_{\rm red} = 0.69$, but
the coefficients describing the radial field are all consistent with zero; in
any case, the remaining model parameters are all identical to within the quoted
errors with those given here.

\begin{figure}
\includegraphics[width=8.5cm]{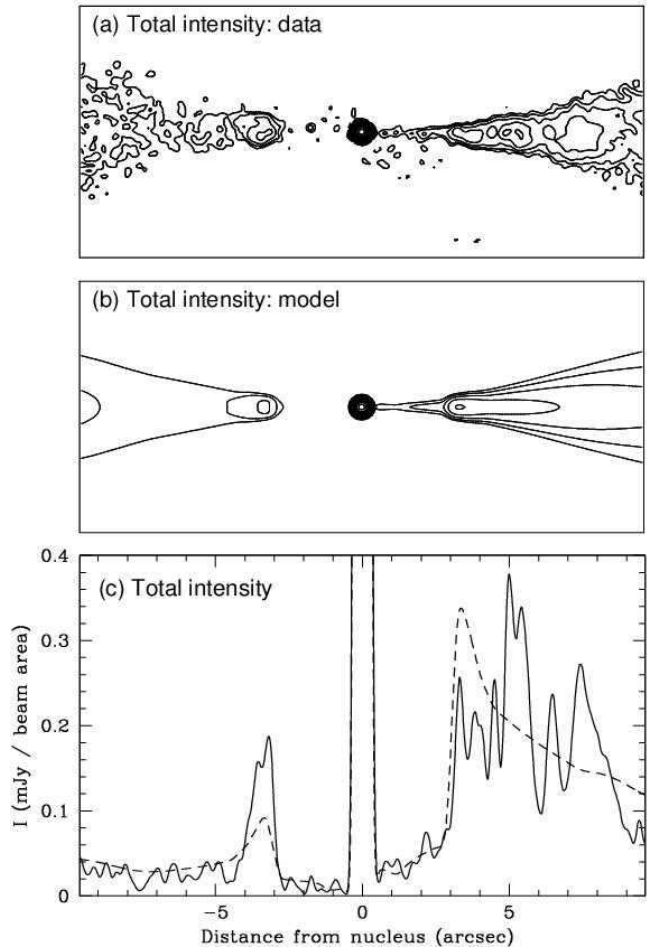}
\caption{Comparison between data and model at a resolution of 0.25\,arcsec. The
  displayed region is $\pm$9.625\,arcsec from the nucleus along the jet axis and
  the scale is indicated at the bottom of the Figure. (a) Observed contours. The
  levels are $-1$, 1, 2, 4, 8, 16, 32, 64, 128, 256, 512, 1024, 2048, 4096
  $\times$ 20\,$\mu$Jy\,beam$^{-1}$. (b) Model contours, with the same levels as
  in panel (a). (c) Profiles of total intensity along the jet axis. Full line:
  data; dashed line: model.  
\label{fig:icomphi}}
\end{figure}

\begin{figure}
\includegraphics[width=8.5cm]{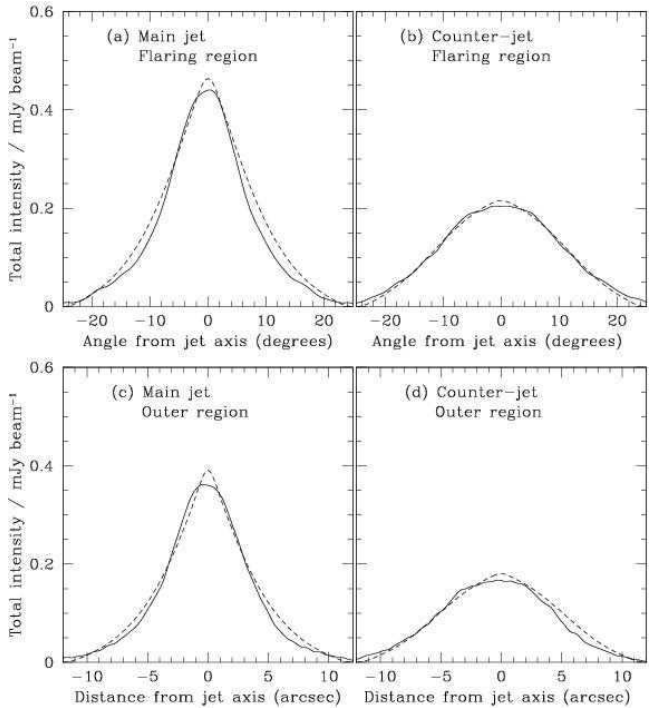}
\caption{Comparison between averaged transverse profiles of total intensity in
  the main and counter-jets at a resolution of 0.75\,arcsec.  (a) main jet, (b)
  counter-jet in the flaring region. The profiles are generated by averaging
  along along radii between 15 and 27 arcsec from the nucleus and plotting
  against angle from the jet axis. (c) main jet, (d) counter-jet in the outer
  region. The profiles are generated by averaging along the jet between 27 and
  40.95\,arcsec from the nucleus and plotting against distance from the jet
  axis. In all panels, the data are represented by full lines and the model by
  dashed lines.
\label{fig:transi}}
\end{figure}

\begin{figure}
\includegraphics[width=6.5cm]{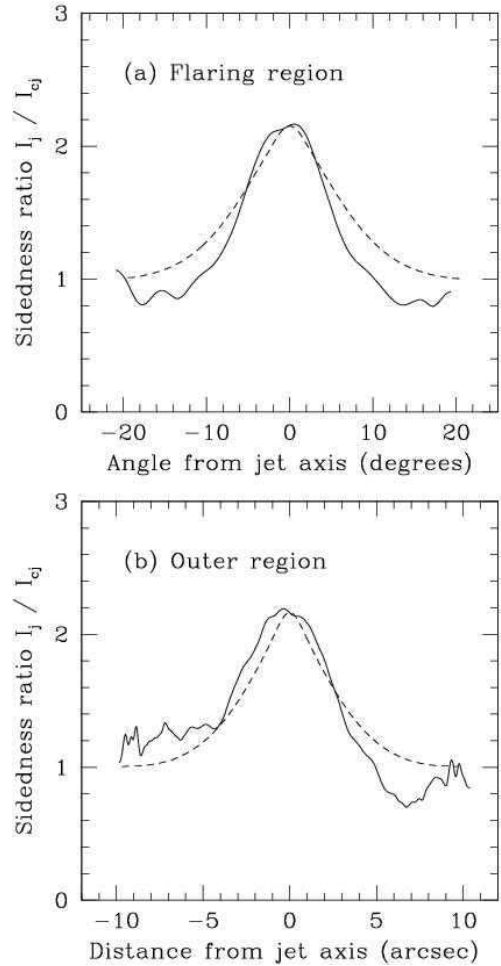}
\caption{Comparison between averaged transverse profiles of jet/counter-jet
  sidedness ratio for data and model at a resolution of 0.75\,arcsec. Full line:
  data; dashed line: model. (a) Averages along radii from the nucleus in the
  flaring region between 15 and 27 arcsec. (b) Average along the jet axis in the
  outer region between 27 and 40.95\,arcsec.
\label{fig:transside}}
\end{figure}

\begin{figure}
\includegraphics[width=8.5cm]{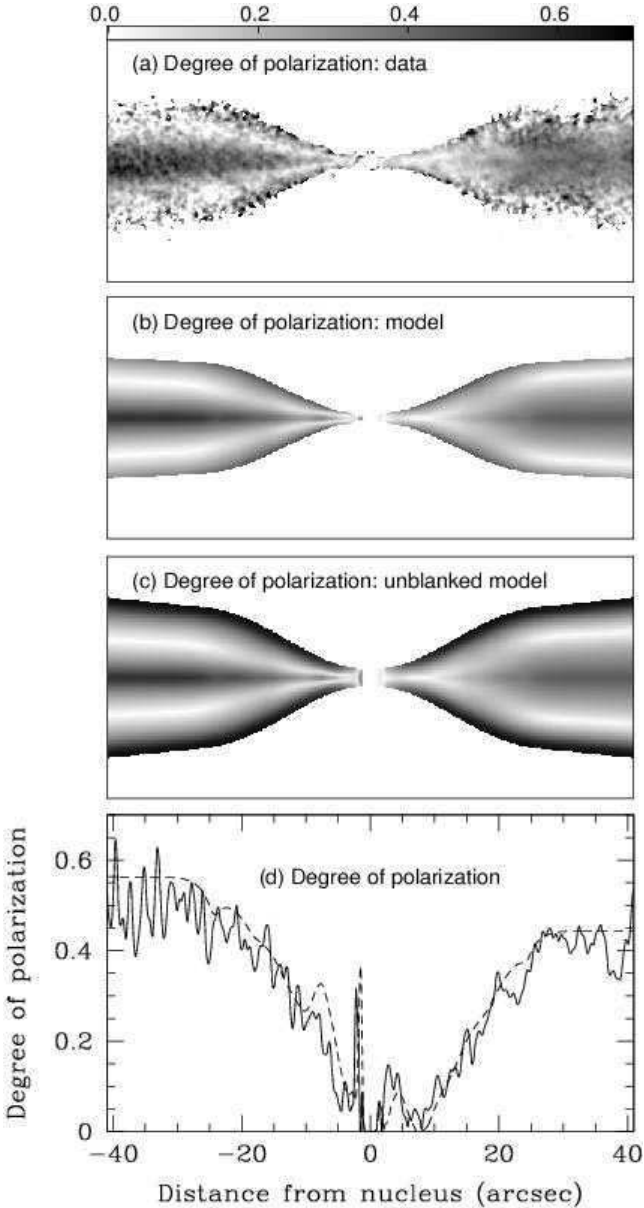}
\caption{Comparison between the degree of polarization, $ p = P/I$, of the data
  and model at a resolution of 0.75\,arcsec. The displayed region is
  $\pm$40.95\,arcsec from the nucleus along the jet axis and the scale is
  indicated at the bottom of the Figure. (a) Grey-scale of observed degree of
  polarization, in the range 0 -- 0.7. Points are only plotted if the total
  intensity $I > 5\sigma_I$ (Table~\ref{tab:noise}). (b) Grey-scale of the model
  degree of polarization in the same range, with the same blanking as in panel
  (a). (c) As in (b), but with no blanking applied. (d) Profiles of $p$ along
  the jet axis. Full line: data (all points have $I > 5\sigma_I$, so no blanking
  has occurred); dashed line: model (also unblanked).
\label{fig:pcomplo}}
\end{figure}

\begin{figure}
\includegraphics[width=8cm]{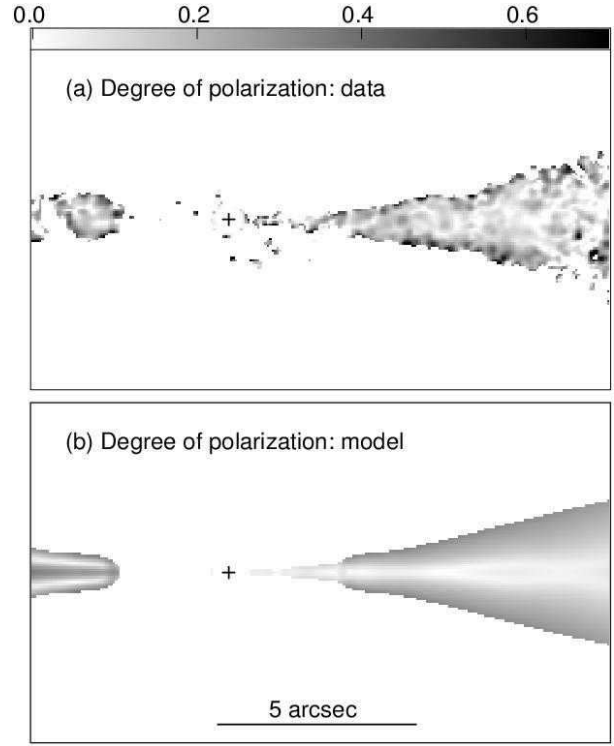}
\caption{Comparison between the degree of polarization, $ p = P/I$, of the data
  and model at a resolution of 0.25\,arcsec. The displayed region is from $-5$
  to $+9.625$\,arcsec from the nucleus along the jet axis and the scale is
  indicated by the labelled bar. The cross shows the position of the core. (a)
  Grey-scale of observed degree of polarization, in the range 0 -- 0.7. Points
  are only plotted if the total intensity $I > 5\sigma_I$
  (Table~\ref{tab:noise}). (b) Grey-scale of the model degree of polarization
  with the same range and blanking as in panel (a).
\label{fig:pcomphi}}
\end{figure}

\begin{figure}
\includegraphics[width=7.8cm]{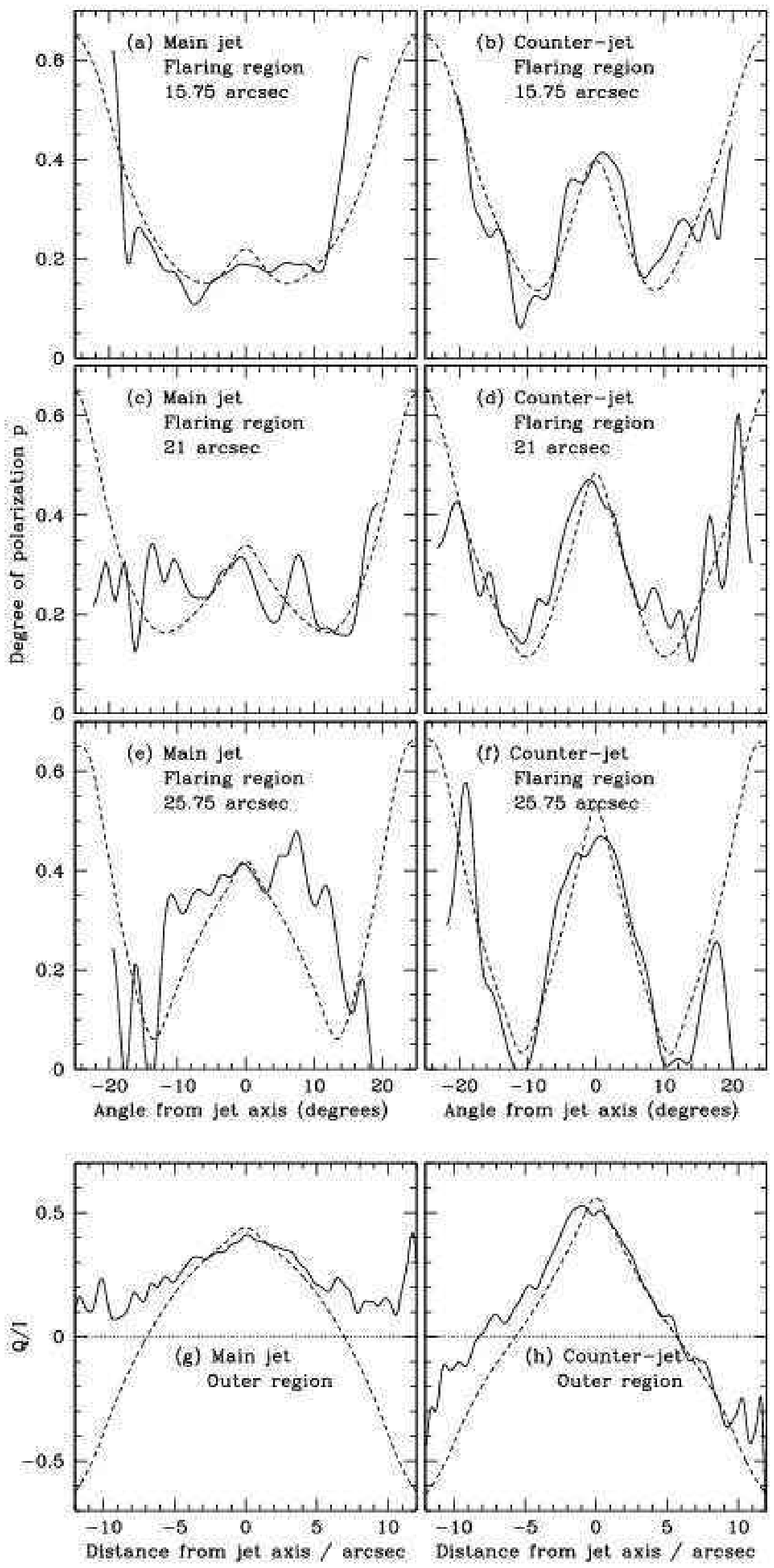}
\caption{Transverse profiles of the degree of polarization for the main jet
  (left) and counter-jet (right) at representative distances from the
  nucleus. The resolution is 0.75\,arcsec FWHM.  In all panels, the data are
  represented by full lines and the model by dashed lines. (a) -- (f): flaring
  region. The profiles are generated by taking scalar averages of $p$ along
  radii from the nucleus and plotting against angle from the jet axis. (a) and
  (b): 15 -- 16.5\,arcsec from the nucleus; (c) and (d): 20.25 -- 21.75\,arcsec;
  (e) and (f) 25 -- 26.5\,arcsec. Panels (g) -- (h): averages of $Q/I$ along the
  jets between 27 and 40.95\,arcsec from the nucleus, plotted against distance
  from the jet axis.
\label{fig:transpol}}
\end{figure}

\section{Comparison between models and data}
\label{results}

\subsection{Total intensity}
\label{Icomp}

The observed and modelled total intensities and jet/counter-jet sidedness ratios
are compared in Figs~\ref{fig:icomplo} --
\ref{fig:transside}. Figs~\ref{fig:icomplo} and \ref{fig:icomphi} show images
and longitudinal profiles at 0.75 and 0.25\,arcsec resolution; averaged
transverse profiles of total intensity and sidedness ratio are plotted in
Figs~\ref{fig:transi} and \ref{fig:transside}.

The following total-intensity features are described well by the model:
\begin{enumerate}
\item The main jet is faint and narrow within $\approx$3\, arcsec of the
  nucleus; such emission as can be seen from the counter-jet is consistent with
  a similar, but fainter brightness distribution  (Fig.~\ref{fig:icomphi}).
\item At 3\,arcsec, there is a sudden increase in brightness on both sides of
  the nucleus.
\item The bright region at the base of the main jet extends to a distance of
  $\approx$13\,arcsec from the nucleus; except for the initial bright knot, the
  counter-jet is much fainter 
  relative to the main jet
  in this region
  (Fig~\ref{fig:icomplo}f).
\item The jet/counter-jet sidedness ratio 
  also
  shows a ridge of nearly constant
  amplitude from 10\,arcsec to the end of the modelled region
  (Figs~\ref{fig:icomplo}d -- f).  The ratio varies from 2.2 on-axis to $\approx
  1$ at the edge of the jet in the outer part of the flaring region and in the
  outer region (Figs~\ref{fig:transside}a, b).
\item The transverse total-intensity profiles from 15\,arcsec to the end of the
  modelled region are accurately reproduced (Fig.~\ref{fig:transi}). 
\end{enumerate}

\subsection{Linear polarization}
\label{Pcomp}

\begin{figure*}
\includegraphics[width=15cm]{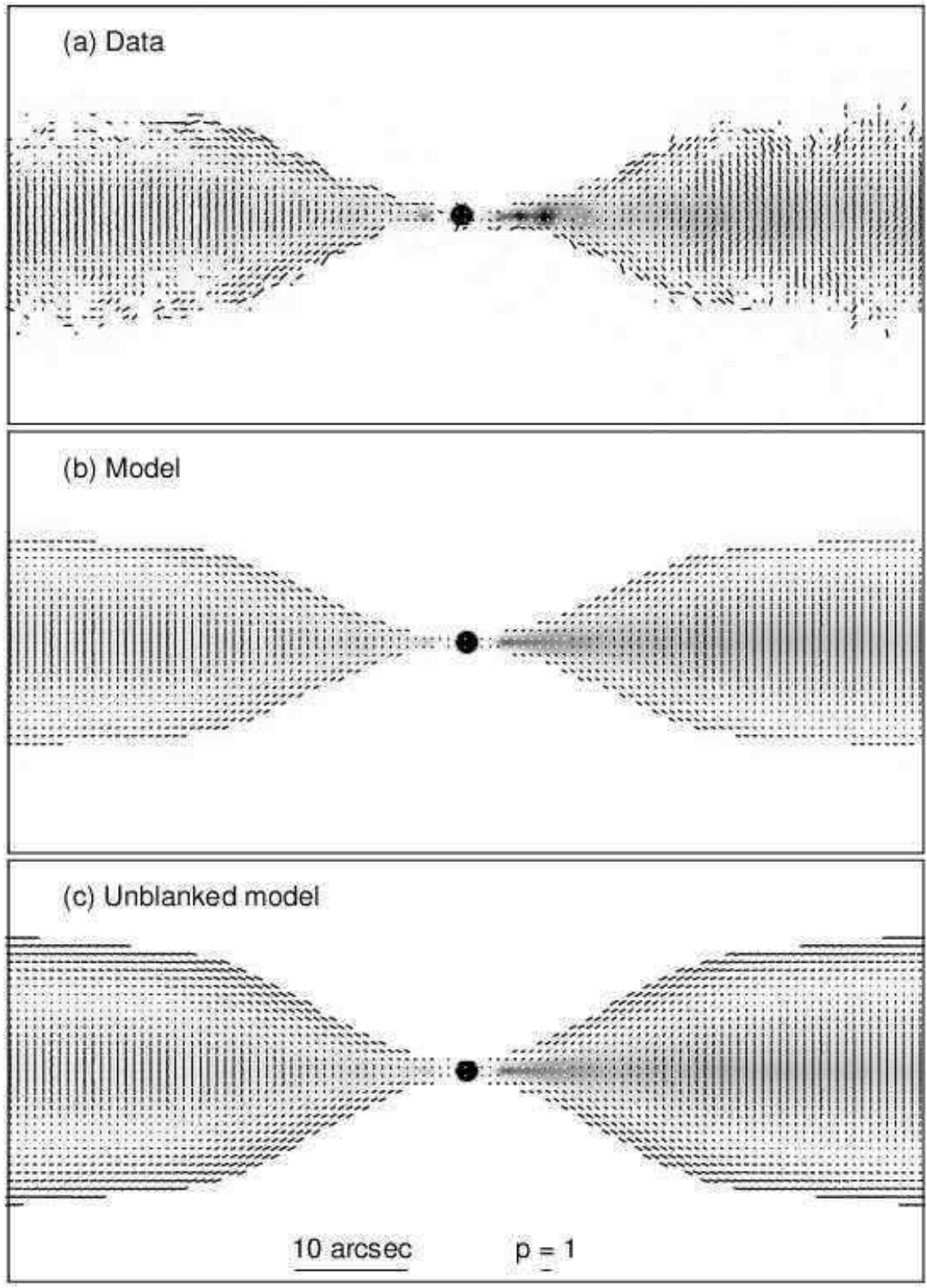}
\caption{Vectors with lengths proportional to the degree of polarization, $p$,
and directions along the apparent magnetic field, superimposed on grey-scales of
total intensity. The resolution is 0.75\,arcsec and vectors are plotted every
0.6\,arcsec. The polarization and angular scales are indicated by the labelled
bars in the lowest panel and the areas plotted are the same as those in
Figs~\ref{fig:icomplo} and \ref{fig:pcomplo}. (a) data; (b) and (c) model. In
panels (a) and (b), vectors are plotted only where $I > 5\sigma_I$
(Table~\ref{tab:noise}); no blanking has been applied in panel (c).
\label{fig:iveclo}}
\end{figure*}

\begin{figure}
\includegraphics[width=8.5cm]{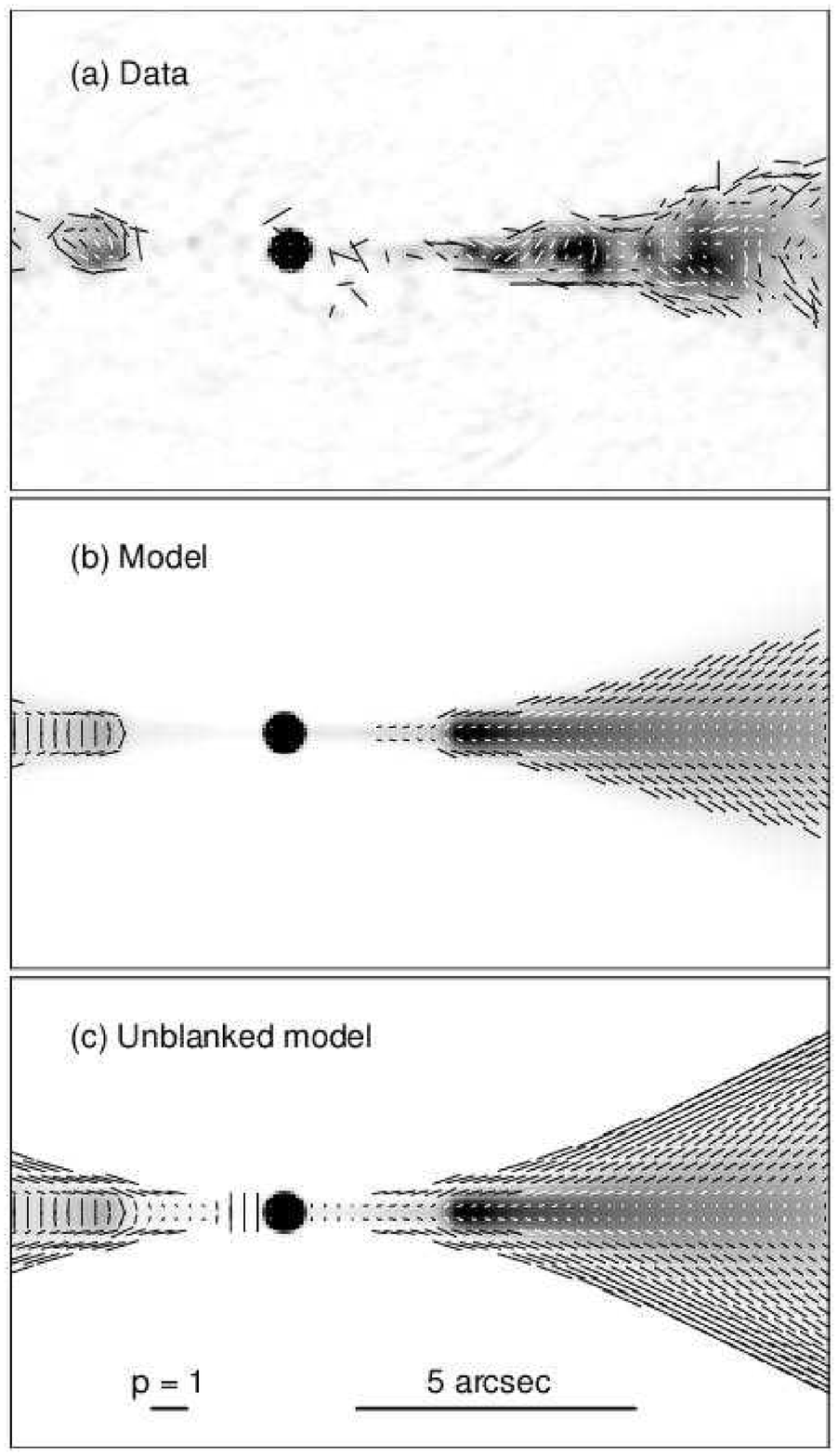}
\caption{Vectors with lengths proportional to the degree of polarization, $p$,
and directions along the apparent magnetic field, superimposed on grey-scales of
total intensity. The resolution is 0.25\,arcsec. The polarization and angular
scales are indicated by the labelled bars in the lowest panel and the areas
plotted are the same as those in Fig.~\ref{fig:pcomphi}. (a) data; (b) and (c)
model. In panels (a) and (b), vectors are plotted only where $I > 5\sigma_I$
(Table~\ref{tab:noise}); no blanking has been applied in panel (c).
\label{fig:ivechi}}
\end{figure}

Grey-scales and longitudinal profiles of the degree of polarization for model
and data at a resolution of 0.75\,arcsec are compared in Fig.~\ref{fig:pcomplo}
and grey-scales of $p$ at 0.25-arcsec resolution are shown in
Fig.~\ref{fig:pcomphi}. The model grey-scales in the former figure are plotted
for $I > 5\sigma_I$, to allow direct comparison with the observations, and
without blanking, to show the polarization of the faint emission predicted at
the edges of the jets.  The observed polarization varies quite rapidly with
distance from the nucleus in the flaring region, so a transverse profile of $p$
averaged over a large range of distances is not useful. We have, instead,
plotted profiles at three representative positions, each averaged over
1.5\,arcsec in distance from the nucleus (Figs~\ref{fig:transpol}a -- f). Since
the position angles vary significantly across the jet, we calculated scalar
averages of $p$.  In the outer region, the transverse $p$ profile is almost
independent of distance, but the polarization at the edges of the jets in the
outer region is difficult to see on the 0.75-arcsec resolution blanked images
profiles, since individual points have low signal-to-noise ratios. In order to
improve the accuracy of the transverse profiles we need to average over $Q$ and
$U$ (as is done by the model fit) rather than the scalar $p$. To do this, we
changed the origin of position angle to be along the jet axis. An apparent field
parallel or orthogonal to the jet axis then appears entirely in the $Q$ Stokes
parameter (and we verified that $U$ was indeed small). We then integrated $Q$
and $I$ along the jet axis and divided to give the profiles of $Q/I$ shown in
Figs~\ref{fig:transpol}(g) and (h). The sign convention is chosen so $Q > 0$ for
a transverse apparent field and $Q < 0$ for a longitudinal one.  Finally, we
plot vectors whose lengths are proportional to $p$ and whose directions are
those of the apparent magnetic field in Figs~\ref{fig:iveclo} (0.75\,arcsec) and
\ref{fig:ivechi} (0.25\,arcsec) with two different blanking levels as in
Fig.~\ref{fig:pcomplo}.  All images of the observed degree
of polarization have first-order corrections for Ricean bias \citep{WK}.

The following features of the polarized brightness distribution are well fitted
by the model.
\begin{enumerate}
\item Both jets show the characteristic pattern of transverse apparent field
  on-axis with parallel field at the edges in the flaring region.
\item The parallel-field edge of the main jet is visible throughout the bright
  region at high resolution, albeit with low signal-to-noise ratio
  (Fig.~\ref{fig:ivechi}). On-axis, the degree of polarization is significantly
  lower. 
\item At high resolution, the apparent field at the base of the counter-jet
  appears to be ``wrapped around'' the bright knot (Fig.~\ref{fig:ivechi}).
\item The longitudinal profile of $p$ (Fig.~\ref{fig:pcomplo}d) is matched well,
  reproducing the significant differences between the two jets.
\item The transition between longitudinal and transverse apparent field (where
  $p = 0$) in the
  main jet is placed correctly at a distance of 8\,arcsec from the nucleus
  (Figs~\ref{fig:iveclo}, \ref{fig:pcomplo}d).
\item In contrast, the counter-jet shows transverse apparent field on-axis even
  very close to the core (Figs~\ref{fig:iveclo}a and c).
\item The asymptotic value of $p$ at distances from the nucleus $\ga$30\,arcsec
  is larger in the counter-jet ($p \approx 0.5$) than in the main jet ($p
  \approx 0.4$). 
\item The ridge of transverse apparent field is broader in the main jet than the
  counter-jet at distances from the nucleus $\ga$25\,arcsec
  (Figs~\ref{fig:pcomplo}a -- c, \ref{fig:transpol}g -- h, \ref{fig:iveclo}). 
\item The evolution of the transverse polarization profiles with distance from
  the nucleus in the flaring region, which differ considerably between the two
  jets, are modelled accurately (Figs~\ref{fig:transpol}a -- f,
  \ref{fig:iveclo}).
\item The apparent field at the edge of the outer region of the counter-jet is
  longitudinal, with a degree of polarization approaching the theoretical
  maximum of $p_0 \approx 0.7$ (Fig.~\ref{fig:transpol}h).
\end{enumerate}

\subsection{Critique}
\label{badfit}

Several features of the total-intensity distribution close to the nucleus are
not well described by the model. The main problem is that the brightness
distributions in both jets show erratic fluctuations causing the local
jet/counter-jet sidedness ratio to increase with distance from the nucleus. Our
models postulate a decelerating flow, so the jet/counter-jet ratio is expected
to decrease monotonically with distance. In fact, the sidedness ratio has a low
value at the position of the first knot in the counter-jet, 3\,arcsec from the
nucleus, compared with its typical value $\approx$10 between 4 and 9 arcsec:
effectively, the first counter-jet knot is roughly a factor of two brighter than
would be expected. The main jet base can be fit more accurately by a model with
less low-velocity emission close to the brightening point, but the counter-jet
knot is then eliminated completely and the overall $\chi^2$ is increased.  The
model for the main jet base also predicts slightly too narrow a brightness
distribution (Fig.~\ref{fig:icomphi}a and b): the fit to the longitudinal $I$
profile of the main jet at high resolution (Fig.~\ref{fig:icomphi}c) is
reasonable, but at lower resolution, where the jet is partially resolved, it
appears worse (Fig.~\ref{fig:icomplo}c). As in all of the other sources we have
modelled, 3C\,296 shows deviations from axisymmetry and erratic intensity
variations in the bright region (Figs~\ref{fig:ihires}b and \ref{fig:icomphi}):
we cannot fit these.

\begin{figure}
\includegraphics[width=7.5cm]{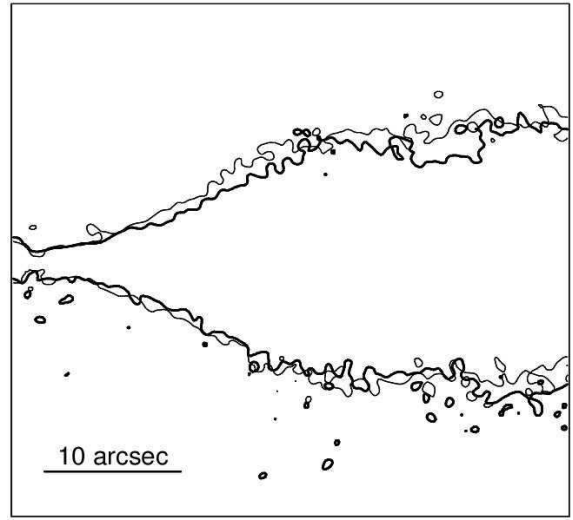}
\caption{Contours of the 20\,$\mu$Jy\,beam$^{-1}$ isophote for the main jet
  (heavy line) and the counter-jet (light line) rotated by 180$^\circ$ about the
  core. This figure illustrates that the counter-jet is slightly wider than the
  main jet in the flaring region. In the outer region, the two isophotes match
  well on one side, but not on the other (primarily because of a lack of
  symmetry in the main jet).
\label{fig:isophote}}
\end{figure}

The counter-jet is slightly wider than the main jet in the flaring region:
Fig.~\ref{fig:isophote} shows a comparison of the 20\,$\mu$Jy\,beam$^{-1}$
isophotes at 0.75\,arcsec resolution. An inevitable consequence is that $I_{\rm
j}/I_{\rm cj} < 1$ at the edges of the jet. This is shown most clearly in the
averaged transverse sidedness profiles of Fig.~\ref{fig:transside}. In the
flaring region, averaging between 15 and 27\,arcsec from the nucleus, the
minimum value of $I_{\rm j}/I_{\rm cj} \approx 0.8$ on both sides.  An
intrinsically symmetric, outflowing, relativistic model cannot generate
sidedness ratios $<$1, so there must be some intrinsic or environmental
asymmetry.  The outer region has an asymmetric transverse sidedness profile,
with $I_{\rm j}/I_{\rm cj} < 1$ on one edge only; this appears to be caused by
deviations from axisymmetry in the main jet (Figs~\ref{fig:icomplo}a and
\ref{fig:isophote}).

Although the qualitative difference in width of the transverse apparent field
region between the main and counter-jets is reproduced correctly, the model
profile is still narrower than the observed one in the outer region of the main
jet. The signal-to-noise ratio is low at the edges of the jets, and the model
fits should not be taken too seriously there, but the averaged profiles of $Q/I$
shown in Fig.~\ref{fig:transpol}(g) suggest that there is a real problem: the
longitudinal apparent field predicted by the model at the edges of the main jet
(an inevitable consequence of the absence of a radial magnetic-field component)
is not seen. In contrast, the corresponding profiles in the counter-jet are
accurately reproduced (Fig.~\ref{fig:transpol}h).  Between $\approx$20 and
60\,arcsec from the nucleus, there is significantly more lobe emission at
8.5\,GHz around the main jet than the counter-jet (Fig.~\ref{fig:ivecdeep}b), so
we have examined the possibility that superposition of unrelated lobe emission
causes the discrepancy between model and data. The diffuse emission around the
main jet has an apparent field transverse to the jet axis, as required to account
for the discrepancy, but is not bright enough to produce the
observed effect.  In order to test whether it is bright enough to affect the
profile in Fig.~\ref{fig:transpol}(g), we fit and subtracted linear baselines
from the $Q$ and $I$ profiles for the jet and counter-jet, using regions between
15 and 20\,arcsec from the axis as representative of the lobe emission. The
subtracted profiles are indistinguishable from those plotted in
Fig.~\ref{fig:transpol}(g) and (h) within 12\,arcsec of the jet axis. Superposed
lobe emission cannot, therefore, be responsible for the discrepancy unless it is
significantly enhanced in the vicinity of the jet.
\footnote{There is significantly more confusing lobe emission at 1.5\,GHz, as
indicated by the steep-spectrum rim observed at 1.5-arcsec resolution
(Fig.~\ref{fig:spec}b).}

\begin{figure}
\includegraphics[width=8.2cm]{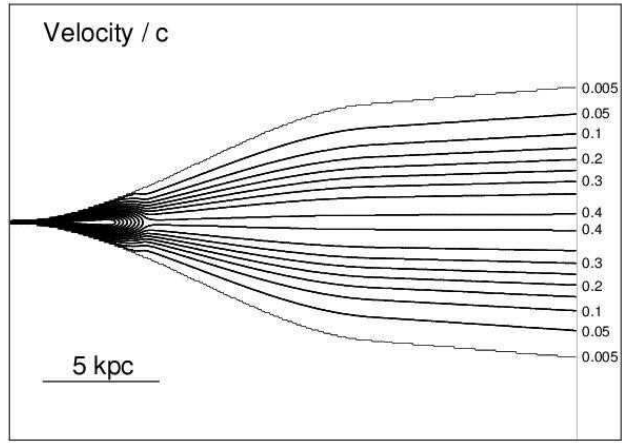}
\caption{Contours of the model velocity field. The thin contour is at $\beta =
  0.005$ and the thick contours are at intervals of
  0.05 in the range $\beta =$ 0.05 -- 0.80.
\label{fig:velcont}}
\end{figure}

\section{Physical parameters}
\label{physical}

\subsection{Summary of parameters}

In this section, all distances in linear units are measured in a plane
containing the jet axis (i.e.\ {\sl not} projected on the sky). The parameters
of the best-fitting model and their approximate uncertainties are given in
Table~\ref{tab:results}.
\begin{table*}
\caption{Fitted parameters and error estimates.\label{tab:results}}
\begin{minipage}{120mm}
\begin{tabular}{llrrr}
\hline
Quantity                 &     Symbol     &~opt~ 
&~min  
&~max\\
\hline
Angle to line of sight (degrees)& $\theta$& 57.9~ &55.9~  & 61.2~ \\
&&&&\\
Geometry                                  &&&&\\
~~Boundary position (kpc)       &  $r_0$  & 15.8~ &14.8~  &16.0~  \\
~~Jet half-opening angle (degrees)&$\xi_0$         
                                            & 4.8~  & 2.9~  & 6.9~  \\
~~Half-width of jet at &$x_0$& 5.2~  & 5.1~  & 5.9~  \\
~~outer boundary (kpc)&&&&\\
&&&&\\
Velocity                                  &&&&\\
~~Boundary positions (kpc)                &&&&\\
~~~~inner                &$\rho_{\rm v_1}$& 4.6~  & 2.3~  & 5.7~  \\
~~~~outer                &$\rho_{\rm v_0}$& 5.7~  & 4.6~  & 6.8~  \\
~~On~$-$~~axis velocities / $c$           &&&&\\
~~~~inner                &   $\beta_1$    & 0.8~  & 0.5~  & $>$0.99 \\
~~~~outer                &   $\beta_0$    & 0.40 & 0.37 & 0.47 \\
~~Fractional velocity at edge of jet      &&&&\\
~~~~inner                &     $v_1$      & 0.1~  & 0.0~  &  0.3~  \\
~~~~outer                &     $v_0$      & 0.04 & 0.02 &  0.08\\
&&&&\\
Emissivity                                &&&&\\
~~Boundary positions (kpc)                &&&&\\
                         &$\rho_{\rm e_{\rm in}}$& 1.8~  & 1.6~  & 2.0~  \\
                         &$\rho_{\rm e_{\rm knot}}$& 2.5~  & 1.9~  & 3.7~  \\ 
                         &$\rho_{\rm e_{\rm out}}$& 8.9~  & 8.4~  & 9.4~  \\
~~On~$-$~~axis emissivity exponents       &&&&\\
                         &     $E_{\rm in}$      & 2.5\footnote{Rough error
  estimates for $E_{\rm in}$ were made by eye (see text).}~   & 2.0    & 3.0      \\
                         &     $E_{\rm knot}$    & 2.1~   & $-0.4$~ & 4.2~   \\
                         &     $E_{\rm mid}$     & 2.8~   & 2.6~     & 3.1~   \\
                         &     $E_{\rm out}$     & 0.99  & 0.88    & 1.12  \\
~~Fractional emissivity at edge of jet    &&&&\\
~~~~inner                &     $e_1$      & 0.8~  & 0.3~  & 2.4~  \\
~~~~outer               &     $e_0$       & 0.12 & 0.10 & 0.19 \\
&&&&\\
B-field                                   &&&&\\
~~Boundary positions (kpc)                &&&&\\
~~~~inner                &$\rho_{\rm B_1}$& 0.0~  & 0.0~  & 2.0~  \\
~~~~outer                &$\rho_{\rm B_0}$& 14.6~ & 12.3~ & 15.9~ \\
~~RMS field ratios                        &&&&\\
~~~~longitudinal/toroidal                 &&&&\\                    
~~~~~~inner region axis  &     $k_1^{\rm axis}$ & 1.7~  & 1.4~  & 2.0~    \\
~~~~~~inner region edge  &     $k_1^{\rm edge}$ & 0.4~  & 0.1~  & 0.6~    \\
~~~~~~outer region axis  &     $k_0^{\rm axis}$ & 0.66 & 0.57 & 0.75   \\
~~~~~~outer region edge  &     $k_0^{\rm edge}$ & 0.0~  & 0.0~  & 0.08   \\
\hline						                    
\end{tabular}
\end{minipage}
\end{table*}

\subsection{Geometry and angle to the line of sight}
\label{results:geom}

The angle to the line of sight from our model fits is $\theta =
58^{+3}_{-2}$\,degree. The axis of the nuclear dust ellipse 
therefore differs from that of the jets by $\approx 15^\circ$ in inclination and
$\approx 31^\circ$ in the plane of the sky \citep{VKdZ05}. The orientation of the dust is such
that the counter-jet would be on the receding side of the galaxy if the dust and
jet axes are even very roughly aligned.  Note, however, that \citet{VKdZ05}
suggest that dust ellipses such as that seen in 3C\,296 are preferentially
aligned with the major axes of the galaxies and statistically unrelated to the jet
orientations.

The outermost isophote of the jet emission is well fitted by our assumed
functional form.  The jets flare to a maximum half-opening angle of 26$^\circ$
at a distance of 8.3\,kpc from the nucleus, thereafter recollimating. The outer
region of conical expansion starts at $r_0$ = 15.8\,kpc from the nucleus and has
a half-opening angle of $\xi_0 \approx 5^\circ$.  The outer envelope of the
model jet emission is shown in Fig.~\ref{fig:profiles}(a).  Note that our models
fit to faint emission outside the 5$\sigma$ isophote of 20\,$\mu$Jy\,beam$^{-1}$
at 0.75-arcsec resolution.

\subsection{Velocity}
\label{results:vel}

\begin{figure*}
\includegraphics[width=16cm]{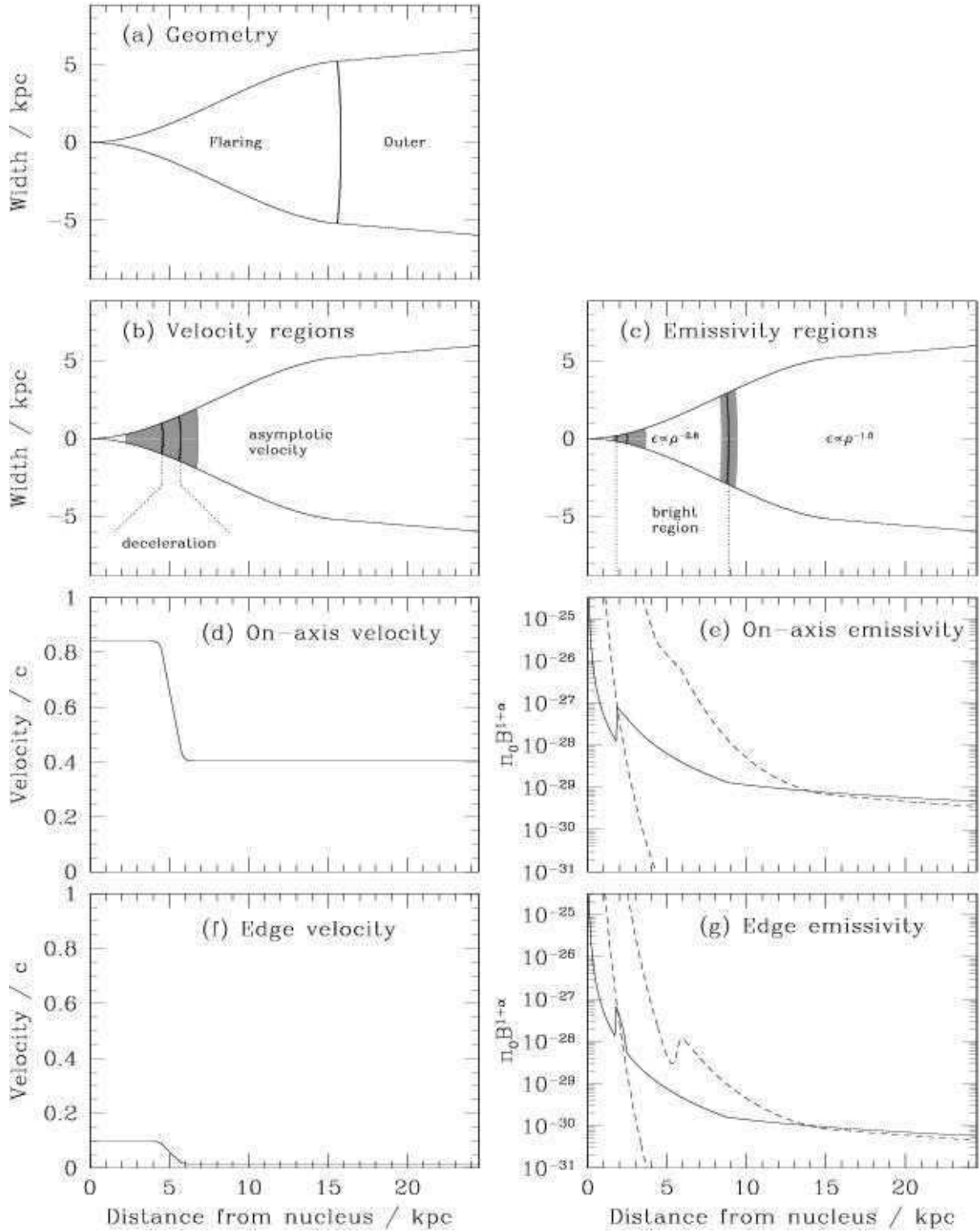}
\caption{(a) The geometry of the jet, showing the flaring and outer regions.  
  (b) and (c):
  sketches showing the relative positions of the boundary surfaces between
  velocity and emissivity regions. The boundaries are defined in
  Table~\ref{tab:param} and their positions for the best fitting model are given
  in Table~\ref{tab:results}. The full curves indicate the boundaries and the
  shaded areas their allowed ranges, also from Table~\ref{tab:results}. (b)
  Velocity. The regions of approximately uniform deceleration and asymptotic
  outer velocity are marked. (c) Emissivity. The region of enhanced emissivity
  between 1.8 and 9\,kpc is indicated.  (d) and (e): profiles of intrinsic 
  parameters along the jet axis in the rest frame. (d) the velocity profile. 
  (e) $n_0 B^{1+\alpha}$
  derived from the emissivity, with $n_0$ and $B$ in SI units. Solid line:
  model; dashed line, adiabatic approximation with the magnetic-field structure
  expected from flux freezing. The latter curve is plotted twice, normalized to
  the model at 1.8 and 14\,kpc from the nucleus, respectively.
  (f) and (g): profiles of intrinsic
  parameters along the jet edge in the rest frame. (f) the velocity,
  (g) the emissivity encoded as in panel (e).
\label{fig:profiles}}
\end{figure*}

Contours of the derived velocity field are shown in Fig.~\ref{fig:velcont}.
The positions of the velocity regions, with their uncertainties, and the
profiles of velocity along the on-axis and edge streamlines are plotted in
Figs~\ref{fig:profiles}(b), (d) and (f).  The initial on-axis velocity is poorly
constrained. The best-fitting value, $\beta_1 = 0.8$, is very uncertain, and the
upper limit is unconstrained by our $\chi^2$ analysis. The reason for this is
that the fractional edge velocity $v_1 = 0.1_{-0.1}^{+0.2}$ is very low: the
emission from the main jet close to the brightening point is dominated by slower
material. Emission from any high-speed component is Doppler-suppressed and
therefore makes very little contribution to the overall $\chi^2$. 
In contrast, after the jets decelerate
the on-axis velocity is well-constrained and consistent
with a constant value of $\beta_0 = 0.40^{+0.07}_{-0.03}$.  The deceleration is
extremely rapid in the best-fitting model (Fig.~\ref{fig:profiles}d), but the
uncertainties in the boundary positions are also large.  The low fractional edge
velocity at larger distances, $v_0 = 0.04^{+0.06}_{-0.02}$, is required to fit
the transverse sidedness-ratio profile (Fig.~\ref{fig:transside}) and is
well-determined. The fact that the sidedness ratio is less than unity in places
implies that there is some intrinsic asymmetry (Section~\ref{badfit}), but the
qualitative conclusion that the edge velocity is low remains valid.

X-ray and ultraviolet emission is detected between 2 and $\approx$10\,arcsec from the
nucleus \citep{Hard05}, corresponding to 1.2 -- 5.9\,kpc in the frame of the
jet. This corresponds to the bright part of the flaring region, up to and
possibly including the rapid deceleration. 

We have evaluated the jet/counter-jet sidedness ratio outside the modelled
region by integrating flux densities in boxes of size 10\,arcsec along the jet
axis and $\pm$15\,arcsec transverse to it. The mean ratio drops from
$\approx$1.7 between 40 and 70\,arcsec from the nucleus to $\approx$1.05 between
70 and 100\,arcsec. It is therefore possible that further deceleration to $\beta
\la 0.1$ occurs at $\approx$40\,kpc in the jet frame. As noted in
Section~\ref{outline-method}, however, the jets bend and lobe contamination becomes
significant, so we cannot model the velocity variation in this region with much
confidence.

\subsection{Emissivity}
\label{results:emiss}

The positions of the emissivity regions, with their uncertainties, and the
profiles of emissivity along the on-axis and edge streamlines are plotted in
Figs~\ref{fig:profiles}(c), (e) and (g). As in all of the 
FR\,I sources we have examined so far,
the on-axis emissivity profile of 3C\,296 is modelled by three power-law
sections separated by short transitions. Close to the nucleus, before the
brightening point at 1.8\,kpc, the power-law index, $E_{\rm in} \approx 2.5 \pm
0.5$. The errors on $E_{\rm in}$ are estimated by eye because the jets are faint
and any change to the index has very little effect on the value of $\chi^2$,
even over a small area close to the nucleus. We model the brightening point as a
discontinuous increase in emissivity by a factor of $1/g \approx 7$. The short
region between 1.8 and 2.5\,kpc is introduced primarily to model the first knots
in the main and counter-jets; its index is poorly constrained because of our
inability to fit the main and counter-jet simultaneously: high values fit the
counter-jet knot well but overestimate the main jet brightness; small values
have the opposite problem. The main section of the bright jet base from 2.5 --
8.9\,kpc is described by a power law of index $E_{\rm mid} = 2.8$. This joins
smoothly onto the outer region ($E_{\rm out} = 1.0$) without the need for a
rapid transition. The transverse emissivity profile is fairly flat at the
brightening point (albeit with large uncertainties) but evolves to a
well-defined Gaussian with a fractional edge emissivity of 0.12 at large
distances.

\begin{figure*}
\includegraphics[width=14cm]{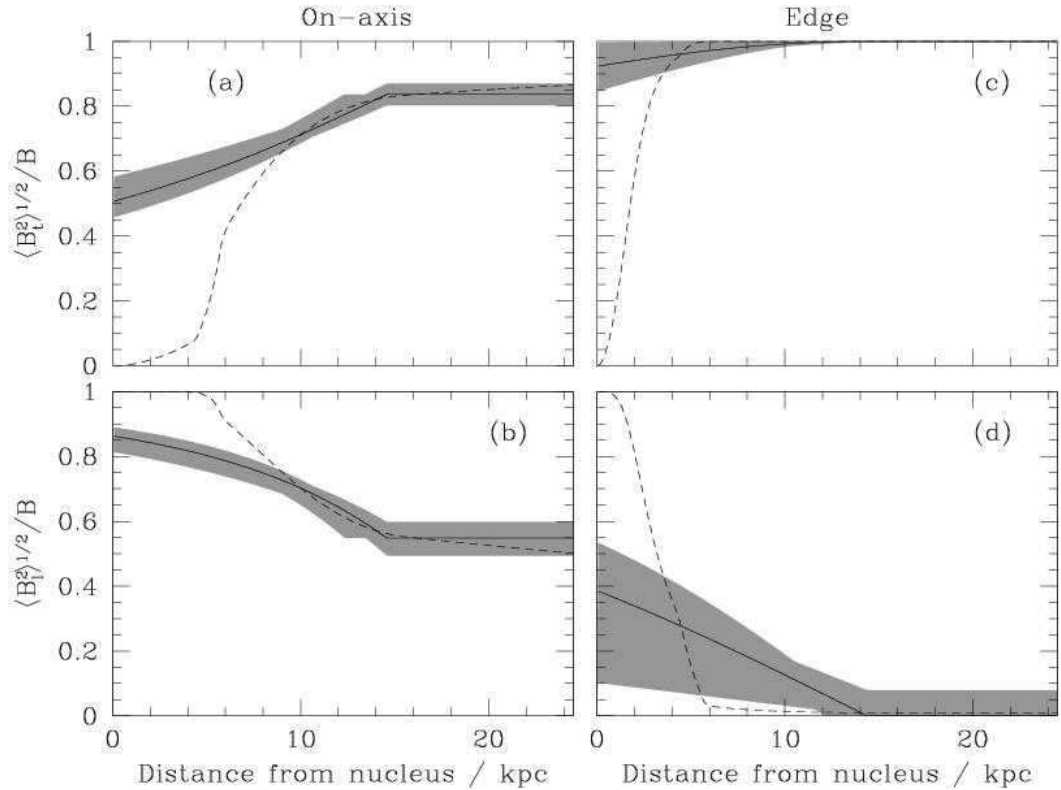}
\caption{Profiles of magnetic-field components along the axis of the jet (panels
  a -- b) and its edge (panels c -- d). The solid lines show the best-fitting
  model, the shaded areas the uncertainties derived from the limits in
  Table~\ref{tab:results} and the dashed lines the profiles expected for a magnetic
  field frozen into the flow.  The profiles for a passively convected field are
  normalized to the free model predictions at a distance of 14\,kpc from the
  nucleus. (a) and (c) toroidal; (b) and (d) longitudinal.
\label{fig:bprofs}}
\end{figure*}

\subsection{Magnetic-field structure}
\label{field-structure}

\begin{figure}
\includegraphics[width=8.2cm]{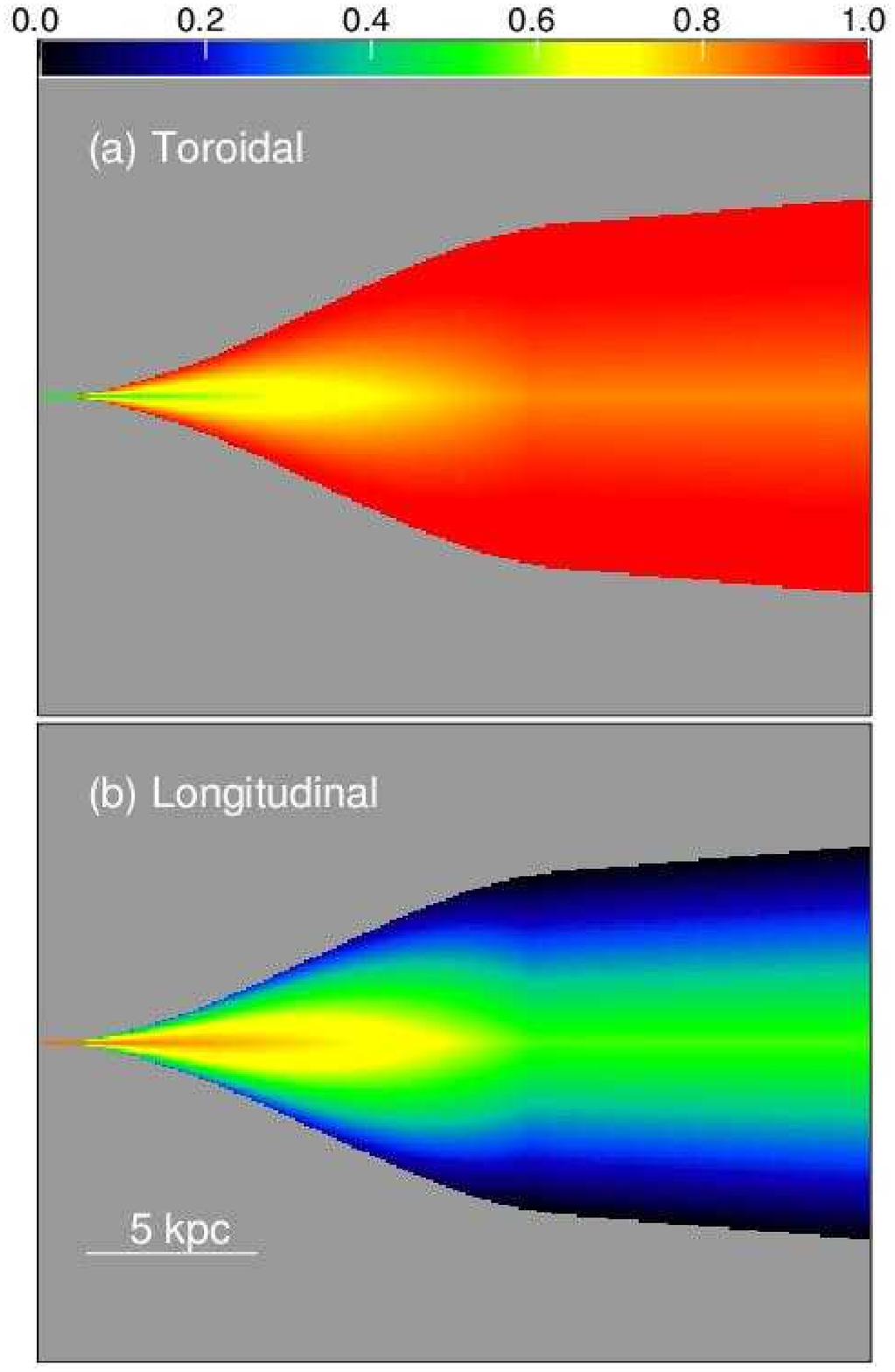}
\caption{False-colour images of the fractional magnetic-field
  components. (a)  toroidal $\langle
  B_t^2 \rangle^{1/2}/B$; (b) longitudinal $\langle B_l^2 \rangle^{1/2}/B$. $B =
  \langle B_t^2+B_l^2 \rangle^{1/2}$.
\label{fig:bcol}}
\end{figure}

The inferred magnetic field components are shown as profiles along the on-axis
and edge streamlines in Fig.~\ref{fig:bprofs} and as false-colour plots in
Fig.~\ref{fig:bcol}. In the former figure, the shaded areas define the region
which the profile could occupy if any one of the six free parameters defining it
is varied up to the error quoted in in Table~\ref{tab:results}.  We were able to
fit the polarization structure in 3C\,296 accurately using a model with only
toroidal and longitudinal magnetic-field components.  The best fitting model has
field-component ratios which vary from the nucleus out to a distance of
14.6\,kpc, consistent with the boundary between flaring and outer regions, and
thereafter remain constant.  The field at the edge of the jet is almost purely
toroidal everywhere (Figs~\ref{fig:bprofs}c, \ref{fig:bcol}a), although there is
a marginally significant longitudinal component close to the nucleus
(Figs~\ref{fig:bprofs}d, \ref{fig:bcol}b). On-axis, in contrast, the
longitudinal field is initially dominant, but decreases over the flaring region
to a smaller, but still significant value at larger distances
(Figs~\ref{fig:bprofs}b, \ref{fig:bcol}b).

We have described the magnetic field structure as disordered on small scales,
but anisotropic. A field of this type in an axisymmetric model always generates
symmetrical transverse profiles of total intensity, degree of polarization and
apparent field position angle. In contrast, a globally-ordered helical field,
unless observed at 90$^\circ$ to the line of sight in its rest frame, will
always generate asymmetric profiles \citep{Laing81,LCB}. The observed profiles
in 3C\,296 (Figs~\ref{fig:transi}, \ref{fig:transpol}) are very symmetrical,
even in regions where the longitudinal and transverse field components have
comparable amplitudes, so we conclude that a simple, globally-ordered field
configuration cannot fit the data. Our calculations would, however, be unchanged
if one of the two field components were vector-ordered; in particular, a
configuration in which the toroidal component is ordered but the longitudinal
one is not would be entirely consistent with our results.
  
The polarization in the high-signal regions closer to the jet axis requires a
mixture of longitudinal and toroidal field components, implying in turn that $p
\approx p_0$ at the edge of the jet with an apparent field along the jet axis.
Our models include this emission, but at a level below the blanking criteria in
total intensity and linear polarization at 0.75-arcsec resolution
(Fig.~\ref{fig:pcomplo}c). We demonstrated in Section~\ref{badfit} that this
emission is indeed present and that it has the correct polarization in the
counter-jet. In the main jet, contamination by surrounding lobe emission
confuses the issue, but there is a real discrepancy (Fig.\ref{fig:transpol}g):
the observed apparent field remains transverse at the edges. The difference
between the main and counter-jets cannot be an effect of aberration as we infer
velocities $\beta \ll 1$ in these regions.  There must, therefore, be a
significant radial field component at the edge of the main jet which is not
present at the corresponding location in the counter-jet.  Although the
low-resolution 8.5-GHz image shown in Fig.~\ref{fig:ivecdeep} gives the
impression that the jet might propagate within its associated lobe while the
counter-jet is in direct contact with interstellar medium, the L-band images
(Fig.~\ref{fig:ifull}) show diffuse emission surrounding {\em both} jets, so
this is unlikely to be the case.

\subsection{Flux freezing and adiabatic models}
\label{ffad}

Given the assumption of flux freezing in a jet without a
transverse velocity gradient, the magnetic field components evolve according to:
\begin{eqnarray*}
B_t &\propto& (x\beta\Gamma)^{-1} \\
B_l &\propto& x^{-2}              \\
\end{eqnarray*}
in the quasi-one-dimensional approximation, where $x$ is the radius of the jet
\citep{Baum97}. The field-component evolution predicted by these equations is
shown by the dashed lines in Fig.~\ref{fig:bprofs}, arbitrarily normalised to
the model values at a distance of 14\,kpc from the nucleus. The evolution of the
longitudinal/toroidal ratio towards a toroidally-dominated configuration is
qualitatively as expected for flux-freezing in an expanding and decelerating
flow, but the model shows a much slower decrease. In the absence of any radial
field, the ratio should not be affected by shear in the laminar velocity field
we have assumed, but the field evolution is clearly more complex than such a
simple picture would predict. A small radial field component could still be
present, and this would allow the growth of longitudinal field via the strong
shear that we infer in 3C\,296. We will investigate this process using the
axisymmetric adiabatic models developed by \citet{LB04}.  A second
possibility is that the flow is turbulent in the flaring region. Two-dimensional
turbulence (with no velocity component in the radial direction) would lead to
amplification of the longitudinal component without generating a large radial
component, as required.  After the jets recollimate, the model field ratios have
constant values on a given streamline, consistent with the variation expected
from flux freezing in a very slowly expanding flow
(Fig.~\ref{fig:bprofs}). Note, however, the problems in fitting the main jet
polarization at large distances from the axis (Section~\ref{field-structure}).

If the radiating electrons suffer only adiabatic losses, the emissivity is:
\begin{eqnarray*}
\epsilon \propto (x^2\beta\Gamma)^{-(1+2\alpha/3)}B^{1+\alpha}
\end{eqnarray*}
in the quasi-one-dimensional approximation \citep{Baum97,LB04}.  $B$ can be
expressed in terms of the parallel-field fraction $f = \langle B_l^2 \rangle
^{1/2}/B$ and the radius $\bar{x}$, velocity $\bar{\beta}$ and Lorentz factor
$\bar{\Gamma}$ at some starting location using equation 8 of \citet{LB04}:
\begin{eqnarray*}
B \propto \left[ f^2\left(\frac{\bar{x}}{x}\right)^4
+ (1 - f^2)\left(\frac{\bar{\Gamma}\bar{\beta}\bar{x}}{\Gamma\beta
x}\right)^2\right ]^{1/2}
\end{eqnarray*}
The resulting profiles for the on-axis and edge streamlines are plotted as
dashed lines in Fig.~\ref{fig:profiles}(e) and (g), normalized to the model values at
the brightening point (1.8\,kpc) and 14\,kpc. Their slopes are grossly inconsistent
with the model emissivity in the flaring region, but match well beyond $\sim$14\,kpc
-- essentially where the jet recollimates at $r_0$ = 15.8\,kpc. 

\subsection{Arcs}
\label{arcs}

\begin{figure*}
\includegraphics[width=17cm]{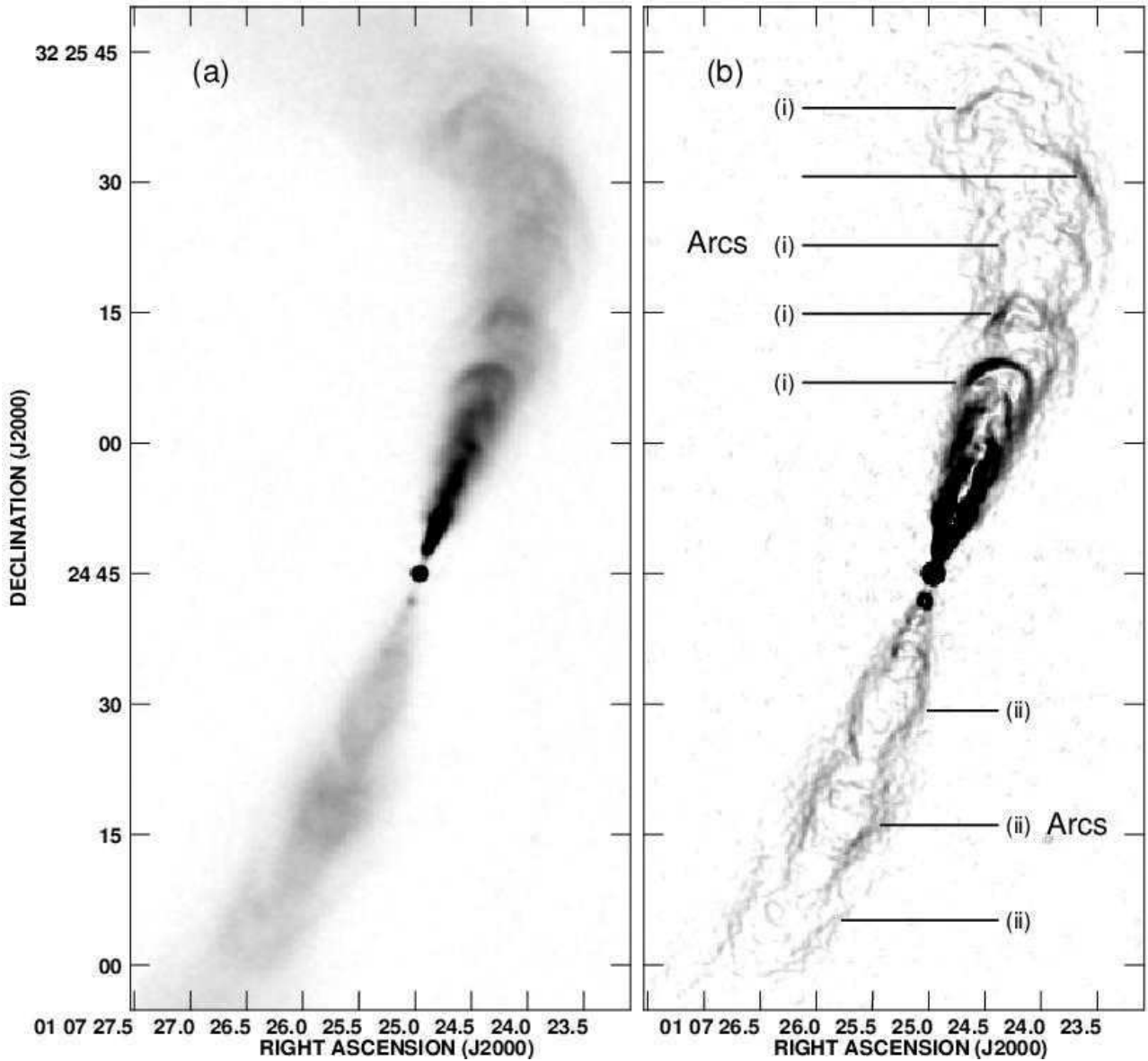}
\caption{8.5-GHz images of the jets in the FR\, radio galaxy 3C\,31 (LB) with
  grey-scale ranges chosen to emphasise the arc structures, as in the equivalent
  plots for 3C\,296 (Fig.~\ref{fig:i1.5}). (a) Total intensity. (b)
  Sobel-filtered $I$. The resolution is 0.75\,arcsec FWHM. The arcs are labelled
  with their types as defined in Section~\ref{arcs} (except for the 
  fourth from the nucleus in the main jet, which appears to be associated with
  the bend in the jet and is clearly not axisymmetric).\label{fig:3c31}}
\end{figure*}

As pointed out in Section~\ref{images}, there are discrete, narrow features
(``arcs'') in the brightness distributions of both jets in 3C\,296. Strikingly
similar structures are seen in the jets of 3C\,31 (LB), as shown in
Fig.~\ref{fig:3c31}.  In both sources, the arcs have systematically different
shapes in the main and counter-jets: centre-brightened arcs with well-defined
intensity gradients near the jet axis [type (i)] are found in the main jets of
both sources and in the outer counter-jet of 3C\,296, while more elongated arcs
with well-defined intensity gradients near the jet edges [type (ii)] are found
in both counter-jets.  We will present models of the arc structures elsewhere:
here we demonstrate that relativistic aberration can plausibly account for the
systematic difference.

Suppose that an arc is a thin, axisymmetric shell of enhanced emissivity,
concave towards the nucleus, and that at any point its speed is roughly that of
the local average flow. Close to the centre-line of the jet, the shell can be
approximated as a planar sheet of material orthogonal to the axis in the
observed frame. Aberration causes a moving object to appear at the angle to the
line of sight appropriate to the rest frame of the flow
\citep{Penrose,Terrell}. For the the normals to the sheets in the main and
counter-jets these angles are $\theta_{\rm j}^\prime$ and $\theta_{\rm
cj}^\prime$, where:
\begin{eqnarray*}
\sin\theta_{\rm j}^\prime  & = & \sin\theta [\Gamma(1-\beta\cos\theta)]^{-1} \\
\sin\theta_{\rm cj}^\prime & = & \sin\theta [\Gamma(1+\beta\cos\theta)]^{-1} \\
\end{eqnarray*}
just as in Section~\ref{outline-method}.

\begin{figure}
\includegraphics[width=8.5cm]{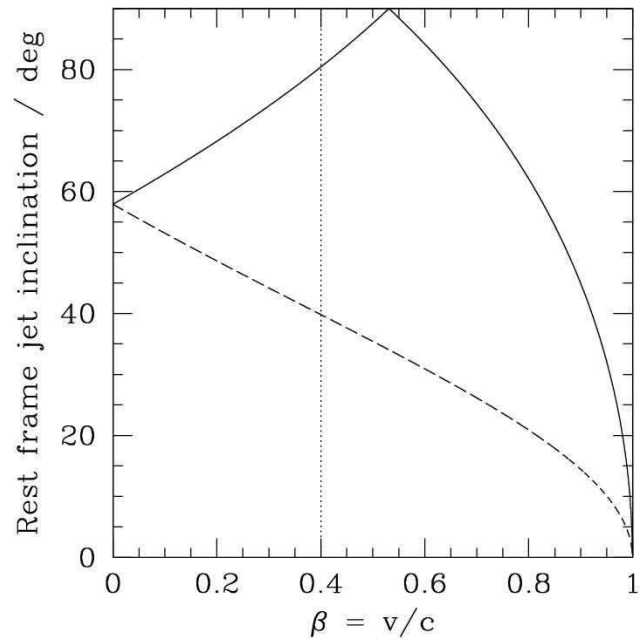}
\caption{The angle to the line of sight in the fluid rest frame for a vector
  parallel to the flow in the observed frame plotted against $\beta = v/c$. The
  angle between the jet axis and the line of sight is $\theta = 58^\circ$, as we
  infer for 3C\,296. Full line: main jet; dashed line: counter-jet. The vertical
  dotted line marks the on-axis velocity of $\beta = 0.4$ we derive for the
  outer parts of the modelled region in 3C\,296. \label{fig:aberrate}}
\end{figure}

In 3C\,296, the arcs are visible as far as $\approx$90\,arcsec from the nucleus
-- outside the region we model -- so we do not have a direct measure of the
average velocity field in their vicinity. Given that our model indicates a
roughly constant velocity on-axis from 10 -- 40\,arcsec from the nucleus we will
assume for the purposes of a rough estimate that the parameters at the outer
edge of the model approximate the average flow parameters near the arcs between
40 and 70\,arcsec, where the mean jet/counter-jet sidedness ratio remains
significantly larger than unity.  In Fig.~\ref{fig:aberrate} we plot
$\theta_{\rm j}^\prime$ and $\theta_{\rm cj}^\prime$ against $\beta$ for $\theta
= 58^\circ$ (Table~\ref{tab:results}).  For a velocity $\beta = \beta_0 = 0.40$,
the angles to the line of sight in the fluid rest frame are $\theta_{\rm j}^\prime
= 80^\circ$ and $\theta_{\rm cj}^\prime = 40^\circ$ (the main jet is close to
the maximum boost condition $\beta = \cos\theta$ which corresponds to edge-on
emission in the rest frame).  A sheet of enhanced emissivity which is normal to
the flow on the jet axis is therefore observed nearly edge-on in the main jet,
appearing narrow, with a high brightness gradient. In the counter-jet, on the
other hand, the same sheet would appear to be significantly rotated about a line
perpendicular to the jet axis in the plane of the sky and therefore less
prominent both in intensity and in brightness gradient.  Close to the edges of
the jets, however, the model flow velocities are essentially zero, so the
effects of aberration should be negligible. If the shells are roughly tangential
to the surface, then we expect to see narrow brightness enhancements orientated
parallel to the edges in both jets.

This picture provides a reasonable qualitative description of the differences in
arc structure for the main and counter-jets in 3C\,296 within
$\approx$70\,arcsec of the nucleus (Fig.~\ref{fig:i1.5}). The evolution of the
mean jet/counter-jet ratio suggests that the jets become sub-relativistic at
larger distances (Section~\ref{results:vel}) and this may explain why we see
type (i) arcs in both jets between 70 and 90\,arcsec.

The hypothesis that the main difference in appearance of the arcs
between the main and counter-jets is an effect of differential relativistic
aberration also allows us to predict that differences should be
observed between main and counter-jet arcs in other FR\,I sources if they are
orientated at moderate angles to the line of sight. This explains why we
observe systematic differences in the arcs of 3C\,31, which has similar model
parameters to 3C\,296 (Fig.~\ref{fig:3c31}, LB), but we would not expect
differences between the main and counter-jet arcs if $\theta_{\rm j}^\prime$
and $\theta_{\rm cj}^\prime$ are similar.  The latter condition holds if the
jets are close to the plane of the sky (e.g.\ 3C\,449 and PKS\,1333-33;
\citealt{Fer99,KBE}) or slow (e.g.\ the outer region of B2\,0326+39; CL).
It will therefore be interesting to search for arcs in deep images of these
sources.


\section{Comparison between sources}
\label{sourcecomp}

We have modelled five FR\,I sources: 3C\,31 (LB), B2\,0326+39, B2\,1553+24 (CL),
NGC\,315 (CLBC) and 3C\,296 (this paper).  We will compare and contrast their
properties in detail elsewhere, but in this section we summarize their
similarities and differences and draw attention to a few potential problems with
our approach.

\subsection{Similarities}

The five FR\,I sources we have studied in detail so far all exhibit the
following similarities:

\begin{enumerate}
\item A symmetrical, axisymmetric, decelerating jet model fits the observed
  brightness and polarization distributions and asymmetries well.
\item The outer boundaries of the well-resolved parts of the jets can be divided
  into two main regions: a {\em flaring} region with a rapidly increasing
  expansion rate followed by recollimation and an {\em outer} region with a
  uniform expansion rate.
\item Close to the nucleus, the on-axis velocity is initially consistent with $\beta
  \approx 0.8$.  On-axis, rapid deceleration then occurs over distances of 1 --
  10\,kpc, after which the velocity either stays constant or decreases less
  rapidly. 
\item All except B2\,1553+24 have low emissivity at the base
  of the ``flaring'' region, so even the main jet is initially very faint 
  (this region may be present but unresolved in B2\,1553+24). 
\item The faint region ends at a ``brightening point'' where the emissivity
  profile flattens suddenly or even increases. This, coupled with the rapid
  expansion of the jets, leads to a sudden increase in surface brightness in the
  main jets (not necessarily in the counter-jets, whose emission is Doppler
  suppressed).
\item The brightening point is always {\it before} the start of rapid deceleration,
  which in turn is completed before the jets recollimate.
\item Immediately after the brightening point, all of the jets contain bright
  substructure with complex, non-axisymmetric features which we cannot model in
  detail.
\item Optical/ultraviolet or X-ray emission has been detected in all of the jets
 observed so far with {\em HST} or {\em Chandra} (results for B2\,0326+39 are not yet
 available). High-energy emission is associated with the bright regions of the
 main jets before the onset of rapid deceleration in all cases, and also with the
 faint inner region in 3C\,31.
\item The longitudinal emissivity profile may be represented by three power-law
  segments separated by short transitions. The indices of the power laws
  decrease with distance from the nucleus.
 \item All of the jets are intrinsically centre-brightened, with fractional
  emissivities at their edges $\approx$0.1 -- 0.5 where these are
  well-determined.
\item On average, the largest single magnetic field component is toroidal. The
  longitudinal component is significant close to the nucleus but decreases with
  distance. The radial component is always the smallest of the three.
\item For those sources we have fit with functional forms allowing variation of
  field-component ratios across the jets (3C\,31, NGC\,315 and 3C\,296), we find
  that the toroidal component is stronger relative to the longitudinal component
  at the edge of the jet (this is likely to be true for the other two sources).
\item The evolution of the field component ratios along the jets before they
  recollimate is not consistent with flux freezing in a laminar flow, which
  requires a much more rapid transition from longitudinal to transverse field
  than we infer.  A possible reason for this is two-dimensional turbulence (with
  no radial velocity component) in the flaring regions.  After recollimation,
  flux freezing is consistent with our results.
\item The inferred evolution of the radial/toroidal field ratio (in those
  sources where the radial component is non-zero) is qualitatively inconsistent
  with flux freezing even if shear in a laminar velocity field is included.
\item The quasi-one-dimensional adiabatic approximation (together with flux
  freezing) is grossly inconsistent with the emissivity evolution required by
  our models before the jets decelerate and recollimate. Earlier attempts to
  infer fractional changes in velocity from surface-brightness profiles 
  \citep{Fanti82,Bick84,Bick86,Bick90,Baum97,Bondi00}, therefore required much
  more rapid deceleration than we deduce in these regions.
\item After deceleration and recollimation, the emissivity evolution along the
  jets is described quite well by flux freezing and a quasi-one-dimensional
  adiabatic approximation (our analysis does not cover this region in NGC\,315).
  As relativistic effects are slight, the brightness evolution in the outer
  regions of 3C\,296, B2\,0326+39 and B2\,1553+24 is quite close to the
  constant-speed, perpendicular-field prediction originally derived by
  \citet{Burch79}. Previous estimates of fractional velocity changes derived on
  the adiabatic assumption [references as in (xv), above] are likely to be more
  reliable on these larger scales.
\end{enumerate}

\subsection{Differences}

3C\,296 exhibits the following differences from the other sources:
\begin{enumerate}
\item The transverse velocity profile falls to a low fractional value $\la 0.1$
  at the edge of the jets in 3C\,296, whereas all the other sources are
  consistent with an edge/on-axis velocity ratio $\approx$0.7 everywhere, and
  also with a top-hat velocity profile at the brightening point.  This could be
  related to the differences in lobe structure: as noted earlier, 3C\,296 is the
  only source we have observed with a bridged twin-jet structure. Its jets may
  then be embedded (almost) completely within the lobes rather than propagating
  in direct contact with the interstellar medium of the host galaxy. Modelling
  of other bridged FR\,I sources would then be expected to show similarly low
  edge velocities.  Whether the lower-velocity material is best regarded as part
  of the jets or the lobes (and, indeed, whether this distinction is meaningful)
  remains unclear. 
\item Some support for the idea of an interaction between jets and lobes comes
  from our detection of anomalous (i.e.\ inconsistent with our model
  predictions) polarization at the edges of the main jet in its outer parts.  We
  have argued that this cannot be a simple superposition of unrelated jet and
  lobe emission.
\item Rapid deceleration is complete before the end of the bright
  region in 3C\,296. In contrast, the bright region ends within the deceleration zone for
  B2\,0326+39, B2\,1553+24 and NGC\,315 and the end of the bright region roughly
  coincides with the end of rapid deceleration in 3C\,31. 
\item When deciding where in the jets our hypothesis of intrinsic symmetry
   first becomes inappropriate, we have successfully used the criterion that the 
   jets must remain straight while selecting the region to be analysed. In 
   3C\,296 however there are hints of intrinsic asymmetries on a still
   smaller scale because the counter-jet appears to be slightly wider than the 
   main jet in the flaring region.
\end{enumerate}

Our results are generally consistent with the conclusions of a statistical study
of a larger sample of FR\,I sources with jets \citep{LPdRF}, but the velocity
range at the brightening point derived from the distribution of jet/counter-jet
ratios assuming an isotropic sample has a maximum $\beta_{\rm max} \approx 0.9$
and a minimum in the range $0.3 \ga \beta_{\rm min} \ga 0$. Such a range would
be expected if low edge velocities (as in 3C\,296) are indeed typical of bridged
twin-jet sources, since these form the majority of the sample \citep{PDF96}.


\section{Summary}
\label{summary}

We have made deep, 8.5-GHz images of the nearby FR\,I radio galaxy 3C\,296 with
the VLA at resolutions ranging from 0.25 to 5.5\,arcsec FWHM, revealing new
details of its twin jets. In particular we see several thin, discrete brightness
enhancements (``arcs'') in both jets. These appear to have systematically
different morphologies in the main and counter-jets.  A comparison with
lower-frequency images from archive data shows that the flat-spectrum jets are
surrounded by a broad sheath of steeper-spectrum diffuse emission, possibly formed by
a backflow of radiating plasma which has suffered significant synchrotron losses
as in models of FR\,II sources. The spectral index of the jets initially
flattens slightly with distance from the nucleus (from $\alpha \approx 0.62$ to
$\alpha \approx 0.53$) as in other FR\,I jets. We have also imaged the Faraday
rotation and depolarization over the source. The counter-jet is more depolarized
than the main jet, but the differences in depolarization and Faraday rotation
variance between the two lobes are small. The rms fluctuations in rotation
measure are larger close to the nucleus in both lobes.

We have shown that many features of the synchrotron emission from the inner
$\pm$40\,arcsec of the jets in 3C\,296 can be fit accurately and
self-consistently on the assumption that they are intrinsically symmetrical,
axisymmetric, decelerating, relativistic flows. The functional forms we use for
the model are similar to those developed in our earlier work, as are most of the
derived physical parameters.  In this sense, a strong family resemblance is
emerging between the physical properties that we derive for the outflows in
these FR\,I sources.

\begin{description}
\item [{\bf Geometry}] The jets can be divided into a flaring region, whose radius $x$
is well fitted by the expression $x = a_2 z^2 + a_3 z^3$, where $z$ is the
distance from the nucleus along the axis, and a conical outer region. We infer
an angle to the line of sight of $58^{+3}_{-2}$\,degree.  The boundary between
the two regions is then at 15.8\,kpc from the nucleus.
\item [{\bf Velocity structure}] Where the jets first brighten, their on-axis velocity
is $\beta \approx 0.8$, but this value is poorly constrained. Rapid deceleration
occurs around 5\,kpc to an on-axis velocity of $\beta = 0.40^{+0.07}_{-0.03}$
which thereafter remains constant. 3C\,296 differs from all of the other objects
we have studied in having a very low velocity (essentially consistent with 0) at
the edges of its jets.
\item [{\bf Emissivity}] The longitudinal profile of rest-frame emissivity is modelled
as three principal power-law sections $\epsilon \propto \rho^{-E}$ with indices
of $E = 2.5$ (0 -- 1.8\,kpc), $2.8$ (2.5 -- 8.9\,kpc) and $1.0$ (8.9 --
24.5\,kpc, the end of the modelled region). The first two regions are separated
by a short transition zone in which the emissivity increases discontinuously by
a factor of $\approx$7 and then has a power-law slope $\sim$2.1 (1.8 --
2.5\,kpc; representing the first knots in the main and counter-jets). The
rest-frame emissivity is centre-brightened.
\item [{\bf Magnetic-field configuration}] We model the magnetic field as disordered,
but anisotropic, with toroidal and longitudinal components, but our results are
equally consistent with an ordered toroidal component.  We rule out a
globally-ordered helical field. The ratio of longitudinal to toroidal field
decreases with distance from the nucleus; the toroidal component is also more
dominant at the edges of the jets.
\item [{\bf Flux freezing}] The field-component evolution is qualitatively consistent
with flux freezing, but much slower than expected in the flaring region, where
the adiabatic approximation also fails to fit the emissivity variation. In the
outer region, after the jets decelerate, the adiabatic approximation is a
reasonable fit.
\item [{\bf Arc asymmetry}] If the arcs are narrow shells of enhanced emissivity
moving with the underlying flow, then the differences in their appearance in the
main and counter-jets can be understood as the effect of relativistic
aberration, as required for a fully self-consistent model of the jets.
\end{description}


\section*{Acknowledgments}

JRC acknowledges a research studentship from the UK Particle Physics and
Astronomy Research Council (PPARC) and MJH thanks the Royal Society for a
University Research Fellowship. The National Radio Astronomy Observatory is a
facility of the National Science Foundation operated under cooperative agreement
by Associated Universities, Inc. We thank Paddy Leahy for providing the
self-calibrated uv data from \citet{LP91} and, in his capacity as referee, for
constructive comments.

\appendix
\section{Functional forms for velocity, emissivity and field ordering}
\label{functional-forms}

In Table~\ref{tab:param}, we give for reference the functional forms for the
spatial variations of velocity, emissivity and magnetic-field ordering used in
this paper. They are identical to those used by CLBC except for the elimination
of any radial field component and a slight change to the form of the
longitudinal emissivity profile. We use the streamline coordinate system
$(\rho,s)$ In the outer region, the streamline index $s = \xi/\xi_0$, where
$\xi$ is the angle between the (straight) streamline and the axis. In the
flaring region, the distance of a streamline from the jet axis at a distance $z$
from the nucleus is $x(z,s)  =  a_2(s) z^2 + a_3(s) z^3$, the coefficients
being defined by continuity of $x(s)$ and its first derivative across the
boundary between the outer and flaring regions. The coordinate $\rho$ increases
monotonically along a streamline and is defined by:
\begin{eqnarray*}
\rho & = & \frac{zr_0\sin\xi_0}{r_0\sin\xi_0+x_0(\cos\xi_0s - 1)} \hspace{2.5cm}
\rho < r_{0} \\
\rho & = & \frac{z\sin\xi_0 +(x_0-r_0\sin\xi_0)(1-\cos\xi_0s)}{\sin\xi_0\cos
    \xi_0s} \hspace{1.0cm} \rho \ge r_{0} \\
\end{eqnarray*}
On-axis ($s = 0$), $\rho = z$ in both regions.

\begin{table*}
\caption{Functional forms of the velocity $\beta$, emissivity
$\epsilon$ and longitudinal/toroidal magnetic-field
ratios $k$ in the streamline coordinate system
$(\rho,s)$. Column 4 lists the parameters which may be optimized, for
comparison with Table~\ref{tab:results}.}
\begin{minipage}{160mm}
\begin{center}
\begin{tabular}{llcl}
\hline
&&&\\
Quantity & Functional form & Range & Free parameters \\
&&&\\
\hline
&&&\\
\multicolumn{4}{c}{Velocity field\footnote{Note that the constants $c_1$,
  $c_2$, $c_3$ and $c_4$ 
are defined by the values of the free parameters and the conditions that the 
velocity and acceleration are continuous at the two boundaries.}}\\ 
&&&\\
$\beta(\rho,s)$ &  $\beta_{1}\exp(-s^2\ln v_1)$&&\\
&$ - \left[\frac{\beta_{1}\exp(-s^2\ln v_1) - \beta_{0}\exp(-s^2\ln v_0)}{10}\right]\exp[c_{1}(\rho - \rho_{\rm v_1})]$ 
& $\rho < \rho_{\rm v_1}$ & Distances $\rho_{\rm v_1}$, $\rho_{\rm
  v_0}$ \\
&&&\\
& $c_{2} + c_{3}\rho$ 
& $\rho_{\rm v_1} \le \rho \le \rho_{\rm v_0}$ & Velocities $\beta_1$,
$\beta_0$\\
&&&\\
& $\beta_{0}\exp(-s^2\ln v_0)$&&\\
&$ + \left[\frac{\beta_{1}\exp(-s^2\ln v_1) - \beta_{0}\exp(-s^2\ln v_0)}{10}\right]\exp[c_{4}(\rho_{\rm v_0} - \rho)]$ 
& $\rho > \rho_{\rm v_0}$ & Fractional edge velocities $v_1$, $v_0$\\
&&&\\
&&&\\
\multicolumn{4}{c}{Emissivity\footnote{The constants $d_1$,
and  $d_2$  
are defined by the condition that the emissivity is continuous except at the
inner boundary.}}\\
&&&\\
$\epsilon(\rho,s)$ & $~~~g\left(\frac{\rho}{\rho_{\rm e_{\rm in}}}\right)^{-E_{\rm in}}$ 
& $\rho \le \rho_{\rm e_{\rm in}}$ & Distances $\rho_{\rm e_{\rm in}}$, $\rho_{\rm
  e_{\rm knot}}$, $\rho_{\rm e_{\rm out}}$ \\
&&&\\
&~~~$\left(\frac{\rho}{\rho_{\rm e_{\rm in}}}\right)^{-E_{\rm knot}}
\exp\left[-s^2\ln\left(e_{\rm in} + (e_0 - e_{\rm in})\left(\frac{\rho - \rho_{\rm e_{\rm in}}}{\rho_{\rm e_{\rm knot}} - \rho_{\rm e_{\rm in}}}\right)\right)\right]$ 
& $\rho_{\rm e_{\rm in}} < \rho \le \rho_{\rm e_{\rm knot}}$ & Indices $E_{\rm in}$, $E_{\rm knot}$, $E_{\rm mid}$, $E_{\rm out}$  \\
&&&\\
& $d_1\left(\frac{\rho}{\rho_{\rm e_{\rm knot}}}\right)^{-E_{\rm mid}}
\exp(-s^2\ln e_0)$ 
& $\rho_{\rm e_{\rm knot}} < \rho \le \rho_{\rm e_{\rm out}}$ &Fractional edge emissivities $e_1$, $e_0$\\
&&&\\
&$d_2 \left(\frac{\rho}{\rho_{\rm e_{\rm out}}}\right)^{-E_{\rm out}} \exp(-s^2\ln e_0)$ 
& $\rho > \rho_{\rm e_{\rm out}}$ &Fractional jump at inner boundary, $g$\\
&&&\\
&&&\\
\multicolumn{4}{c}{Longitudinal/toroidal field ratio}\\
&&&\\
$k(\rho,s)$ &  $k_{1}^{\rm axis} + s(k_{1}^{\rm edge}- k_{1}^{\rm axis})$  
& $\rho \le \rho_{\rm B_1}$ & Ratios  $k_1^{\rm edge}$, $k_0^{\rm
  edge}$, $k_1^{\rm axis}$, $k_0^{\rm axis}$ \\
&&&\\
&$k^{\rm axis} + s(k^{\rm edge}- k^{\rm axis})$ 
& $\rho_{\rm B_1} < \rho < \rho_{\rm B_0}$   & Distances $\rho_{\rm B_1}$, $\rho_{\rm B_0}$\\\
&where~~$k^{\rm axis} = k_{1}^{\rm axis} + (k_{0}^{\rm axis} - k_{1}^{\rm axis})\left(\frac{\rho - \rho_{\rm B_1}} {\rho_{\rm B_0} - \rho_{\rm B_1}}\right)$ 
& &\\
&~~~~~~~~~$k^{\rm edge} = k_{1}^{\rm edge} + (k_{0}^{\rm edge} - k_{1}^{\rm edge})\left(\frac{\rho - \rho_{\rm B_1}} {\rho_{\rm B_0} - \rho_{\rm B_1}}\right)$ 
& &\\
&&&\\
& $k_{0}^{\rm axis} + s(k_{0}^{\rm edge}- k_{0}^{\rm axis})$ 
& $\rho \ge \rho_{\rm B_0}$ & \\
&&&\\
\hline
\end{tabular}
\end{center}
\end{minipage}
\label{tab:param}
\end{table*}

\label{lastpage}
\end{document}